\documentclass[%
linenumbers,
superscriptaddress,
 amsmath,amssymb,
 aps, physrev,
]{revtex4-2}

\usepackage{array}
\usepackage[table]{xcolor}

\usepackage{graphicx}
\usepackage{dcolumn}
\usepackage{bm}
\usepackage{hyperref}
\usepackage{color}

\usepackage[separate-uncertainty=true]{siunitx}

\usepackage{cancel}

\usepackage[normalem]{ulem}
\usepackage{xcolor}

\newcommand\dd{\text{d}}

\newcommand\por{\phi}  
\newcommand\permeability{\kappa}
\newcommand\dsdz{\psi}      
\newcommand\dia{a}   
\newcommand\thick{h}
\newcommand\scalefact{S}

\begin{document}


\preprint{APS/123-QED}

\title{\textbf{Transport and removal of a passive tracer in porous media employing surface washing} 
}%

\author{Georgia Ioannou}
\email{Contact author: gi240@cantab.ac.uk}
 \affiliation{ Department of Applied Mathematics and Theoretical Physics, University of Cambridge, Cambridge CB3 0WA, United Kingdom}
\author{Francesco Paolo Cont\`o}
 \affiliation{ Department of Applied Mathematics and Theoretical Physics, University of Cambridge, Cambridge CB3 0WA, United Kingdom}
\author{Merlin A. Etzold}
\affiliation{%
Defence Science and Technology Laboratory, Porton Down, Salisbury,
Wiltshire, United Kingdom
}%
\author{Julien R. Landel}
\affiliation{%
Universite Claude Bernard Lyon 1, Laboratoire de Mécanique des Fluides et d'Acoustique (LMFA), UMR5509, CNRS, Ecole Centrale de Lyon, INSA Lyon, 69622
Villeurbanne, France,
}%
\affiliation{%
Department of Mathematics, University of Manchester, Oxford Road, Manchester, M13 9PL, United Kingdom.
}%
\author{Stuart B. Dalziel}
 \affiliation{ Department of Applied Mathematics and Theoretical Physics, University of Cambridge, Cambridge CB3 0WA, United Kingdom}

\date{\today}

\begin{abstract}
This experimental study investigates the dynamics of surface washing to remove a passive tracer from a porous plate by a gravity-driven liquid film flow. A disodium fluorescein tracer is introduced at the surface of a water-saturated porous plate and allowed to diffuse into the plate for a number of hours before a water film flows over its surface to extract and transport the tracer away. The removal rate of the tracer is monitored quantitatively by using fluorescence measurements to determine the tracer concentration in the effluent of the washing process. These measurements are supplemented by dye-attenuation imaging, which provides some qualitative and semi-quantitative insights about the planar concentration distribution of the tracer integrated across the depth of the porous plate. Our findings reveal a three-stage mass-transport process. First, the tracer present in the surface roughness of the porous plate, in the superficial layer at the interface with the washing film flow, is rapidly removed through advective processes, a stage we refer to as surface flushing. Second, while the tracer patch is transported downstream within the porous material, a slower surface-normal transport process (diffusion enhanced by dispersion) limits the removal of the tracer out of the porous material by the film flow, defining an advection--dispersion stage. Third, as the patch reaches the downstream boundary of the porous plate advective transport processes accelerate its removal in an expulsion stage. A parametric study explores the influence of the characteristics of the washing film, the permeability of the porous plate, the amount and initial spatial extent of the tracer on the porous plate and the initial penetration depth of the tracer, determined by the duration of its diffusion in the porous plate before the washing experiment, on the mass removal rates in the different stages. The normalised mass removal rate increases with steeper substrate inclinations, owing to stronger surface-normal dispersion within the porous medium and advective transport within the surface film. Moreover, the data exhibit self-similar behaviour across different inclination angles when time is scaled by the sine of the inclination angle, which relates to the effective impact of gravity. Increasing the dwell period, defined as the duration between tracer deposition and the beginning of surface washing, allows the tracer to penetrate deeper into the medium and therefore reduces the amount of tracer removed during the first stage, while an increased removal rate is observed in the second stage. Experiments on porous plates with different permeabilities showed that smaller amounts of tracer are removed from lower-permeability plates during the first stage, the advective tracer transport downstream is slower and that the tracer removal rate in the second stage scales with the advection time scale within the porous medium. As the third stage is advection dominated, its duration also scales with the advection time scale within the porous medium. Therefore, we find that the overall cleaning time scales with the advection time scale within the porous medium highlighting the dominant role of advection in this process. These insights offer practical guidance for optimising surface washing protocols for porous systems in industrial and environmental applications.

\end{abstract}
\nolinenumbers

\maketitle


\section{\label{sec:intro}Introduction}
Many materials encountered in everyday life are porous. Examples include concrete, brick, mortar, ceramics, unglazed tiles, plaster, and asphalt \cite{Hu2013,Hall2021,Fusade2019}. Contaminants, such as chemical and radioactive substances as well as biological pathogens that pose significant risks to human health, can penetrate and become absorbed into the pore space of materials, from which they are often difficult to remove \cite{Fitch2003,Settles2006,Boone2007,Gazi2025}. Depending on their physicochemical properties, such contaminants can be broadly classified as short-lived or persistent \citep{Ryan2008}.

Short-lived contaminants tend to degrade or evaporate naturally within minutes or hours due to chemical reactions such as hydrolysis, environmental conditions such as airflow over the surface \cite{Griffiths1999}, and biodegradation \citep{Bartelt-Hunt2006}. Porous media properties, such as wettability and pore connectivity, have been shown to play a critical role in controlling the transport (especially permeation and desorption) of contaminants through the porous network \cite{Jenkins1994}. Transport and physicochemical transformation of contaminants within and outside on the surface of the porous materials have been extensively studied \cite{Westin1998,Willis2012a,Willis2012b}. Experimental and theoretical investigations have shown that evaporation can lead to partial removal of absorbed species, and vapour-phase transport may temporarily increase airborne exposure risks. 

In contrast, persistent contaminants remain within the porous matrix for extended periods (ranging from days to even years) due to their chemical stability and low vapour pressure. These substances pose a risk of chronic exposure, primarily through dermal contact rather than inhalation, as their low volatility limits their partitioning into the gas phase. Persistent contaminants may also re-emerge if the material structure is compromised, either through human activity (e.g., drilling, scraping, or sanding) or natural degradation processes (e.g., cracking, erosion, or material fatigue).

Cleaning and decontamination of porous surfaces is a complex process governed by diffusion, advection, and chemical and biological reaction mechanisms, and is strongly influenced by factors such as pore structure, surface heterogeneity, and the interplay between surface and bulk fluid dynamics \cite{Wang2013,Landel2021,Geng2023}. The persistence of pathogens in porous materials, combined with their heterogeneous distribution across different materials, highlights the practical difficulties and inherent complexity of achieving effective remediation and decontamination \cite{Boone2007}. Additionally, redistribution of the contaminant within a porous medium can also occur during the decontamination process. Despite the importance of surface-washing techniques in industrial and hazard-management emergency-response contexts, the most used decontamination methods remain largely empirical \cite{Fryer2009,Landel2021,Wilson2022}, with only a limited number of fundamental studies addressing the underlying transport mechanisms. Consequently, both the design of industrial cleaning protocols and the development of hazard management strategies would benefit from a deeper physical understanding of the governing dynamics.

In cases where decontamination involves chemical reactions between the contaminant and the decontaminant, or phase-change processes such as evaporation or dissolution, the transport dynamics within porous media are governed by various coupled mechanisms. Previous works have analysed the effect of solubility of the products of a reaction on diffusive--reactive transport dynamics involving immiscible liquid--liquid contaminant--decontaminant systems within porous media \cite{Dalwadi2017}. The evolution of the interface between contaminant and decontaminant has been modelled as a moving boundary, following a Stefan-type formulation, where diffusion and reaction dominate while advection is neglected. Depending on the relative diffusive and reactive timescales, a diffusive boundary layer can form, slowing down the advance of the reactive front and thus the overall decontamination process \cite{Murphy2023}. Additional studies have explored contraction or swelling effects caused by density changes between products and reactants \cite{Geng2023}. Furthermore, theoretical works have used homogenisation techniques to derive macroscopic models that describe pore-scale evaporative or reactive fronts within porous media \cite{Luckins2023,Luckins2024}. 

Although decontamination by surface washing using high-pressure liquid jets and films is a common technique in both everyday life and industry to clean surfaces \cite{Landel2021}, the mechanisms of this process have received comparatively little attention in the scientific literature. A recent study on porous media decontamination was conducted by Maryshev and Klimenko \cite{Maryshev2023}, who investigated numerically the cleaning of a porous medium clogged by fine, immobile, insoluble particles of solute using an external flow. Their findings demonstrated that the cleaning process was accelerated by increasing the external flow flux and increasing the density difference between the solute particles that clog the pores and the cleaning fluid, given that buoyancy-driven forces act in the solute removal direction. Moreover, the cleaning process is decelerated by an increase in the adsorption of the solute particles by the media, which immobilises them. Therefore, it is crucial to understand how advection-dominated transport at the surface and/or below the surface affects the cleaning of porous substrate. In this work, we address the lack of fundamental studies that investigate the coupling between transport in the surface washing film and within the porous media. Specifically, we examine the dynamics of the removal of a tracer from a porous plate subjected to a surface washing film.

Experimentally, flow through porous media is often characterised using tracers, substances that serve as markers and track the fluid movement through the pore spaces to provide insight into the flow dynamics and the properties of the medium. Passive tracer methods have been employed to directly measure groundwater and contaminant fluxes in porous substrates, providing quantitative insight that shows that decontamination efficiency is significantly influenced by spatial and temporal heterogeneities in the advective flow within the porous structure \cite{Hatfield2004}. It has also been shown that the permeability field of a porous medium can be inferred from passive tracer injections under controlled laboratory conditions using X-ray tomographic visualisation \cite{Zhan2000}. In the oil and gas industry, chemical and radioactive tracers are often utilised for assessing fluid flow, evaluating sub-surface porous hydrocarbon reservoirs and study fluid migration mechanisms between wells and drilled fluid pathways \cite{Patidar2022}. 

Techniques for flow visualisation in porous media are often costly and require specialised equipment and methodologies. In this work, we use a simple, affordable and safe technique, readily available in the toolbox of most fluid mechanics laboratories. We employ a water-soluble dye as our tracer and we adapt the light attenuation method \cite{Cenedese1998,Allgayer2012} for a porous medium in order to obtain the depth-integrated space and time resolved in situ tracer distribution across a large planar area of the substrate. This technique can achieve virtually any temporal resolution of relevance to porous media transport, even for high permeability porous media, contrary to X-Ray tomography techniques which are limited to a few seconds in temporal resolution at best \cite{Hasan2020,VanOffenwert2021}. Simultaneously, in this work we measure the concentration of tracer in the effluent from the washing process through fluorometry. We demonstrate that this method provides highly accurate data.

The removal of a tracer from a porous medium by surface washing constitutes a very rich and complex problem space. Both the fluid properties and the porous media characteristics play critical roles. For example, the porous medium may initially be dry or (partially) saturated with a fluid, which immediately introduces miscibility effects between tracer, pore fluid, and washing fluid and wettability effects with the porous medium. Realistic cleaning scenarios often involve highly non-uniform initial tracer distributions, further expanding the parameter space of the problem. Additional complexities arise from chemical reactions between the contaminant and the washing fluid, variations in pore size and tortuosity, heterogeneities in material permeability, the presence of surface roughness, and boundary effects at edges or interfaces. Temperature gradients, evaporation, and phase changes within the porous medium can also influence transport, making even seemingly simple systems highly non-trivial to analyse experimentally, theoretically or numerically \cite{Ladd2021,Yang2024}. Even when focusing on the  simple configuration studied in this work, the removal of a passive tracer with a known initial concentration distribution from an initially water-saturated porous plate with relatively uniform porosity $\por$ and permeability $\permeability$, using a controlled gravity-driven surface washing film, the process is not as simple and intuitive as it may seem. The experiment involves significant practical challenges including the difficulty of accurately monitoring the flow and achieving reproducible results, while the underlying transport processes involve an interplay between advection (induced by gravity on the surface flow and within the porous plate) and diffusion both within the porous substrate and into the fresh washing film from tracer-rich regions of the porous plate. Parameters such as film velocity, porous plate properties, and the initial conditions have a great influence on the observed dynamics. Understanding this process is essential not only for interpreting small-scale experimental results, but also for designing larger scale trials that can be used for developing protocols and effective cleaning and decontamination strategies for porous materials in various contexts.

Details of the experimental method are presented in Sec.~\ref{sec:methods}, including the presentation of the experimental set-up and procedure, the characterisation of the porous plate, and the description of methods employed for data acquisition. In Sec.~\ref{sec:results}, the mechanisms governing the transport of a passive tracer out of the porous matrix are examined and a parametric study investigating the influence of key parameters on mass-transfer dynamics is presented. Finally, the main findings and conclusions are summarised in Sec.~\ref{sec:conslusions}.

\section{\label{sec:methods}Materials and methods}

\begin{figure*}[!ht]
\includegraphics[width=0.85\textwidth]{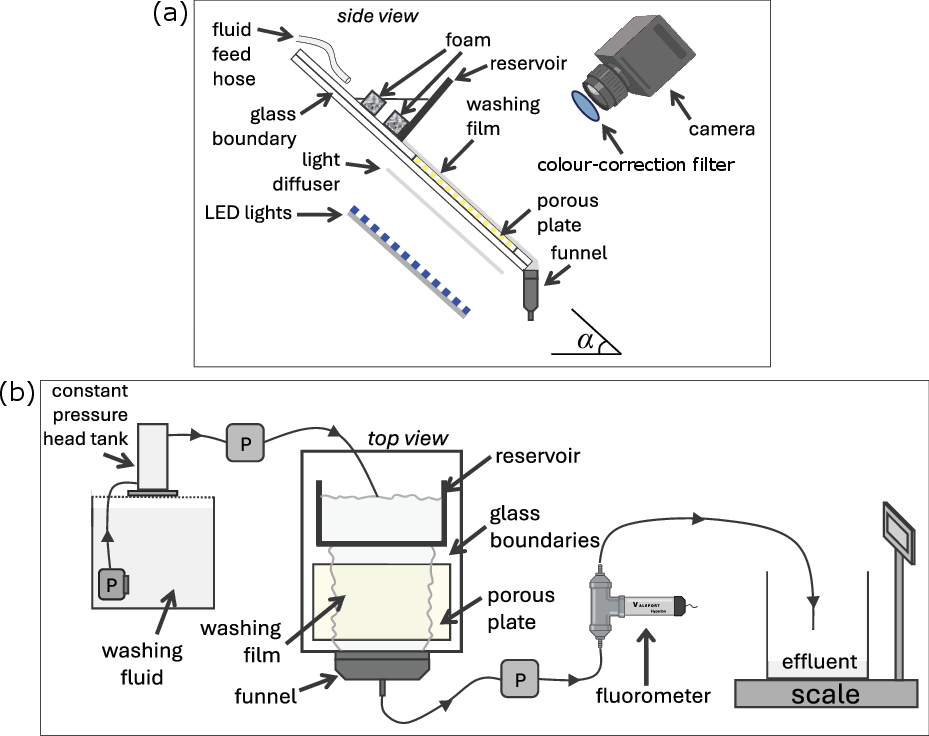}
\caption{\label{fig:setup} Schematics of the experimental setup. (a) Side view of the surface-washing apparatus. (b) The complete experimental setup, including a top view of the washing apparatus. Squares with P represent gear pumps.}
\end{figure*}

\subsection{\label{sec:setup}The experimental setup and procedure}

The experimental setup used in this work is based on a surface washing apparatus adapted from the design originally utilised by \cite{Landel2016} for impermeable surfaces. A schematic representation of the side view of the washing apparatus can be seen in Fig.~\ref{fig:setup}(a). The porous substrate, embedded in a larger glass plate (see Sec.~\ref{sec:plate}), was inclined at an angle $\alpha$ with respect to the horizontal plane during experiments. Disodium fluorescein solutions were used as passive tracers on the porous plate and sodium carbonate/bicarbonate buffered softened tap water as the washing fluid. The film flow was fed from a reservoir system designed to ensure uniformity across the transverse film width and dampen any fluctuations. Specifically, the incoming washing solution passed through a two-level distribution manifold with foam sponge before exiting from the reservoir gate through a \SI{2.25}{\milli \metre} opening set by a micrometer screw controller. At the lower end of the plate, the effluent (the washing film that has passed from the porous plate and contains traces of the fluorescein dye) fell into a funnel. 

A JAI SP-45000M-CXP4-F monochrome video camera (fitted with a Nikkor \SI{60}{\milli \metre} lens) was mounted above the porous plate with its optical axis normal to the surface of the plate. In combination with a panel of blue LEDs (with peak wavelength around 460--470\si{\nano\metre}) paired with a light diffuser underneath the porous plate, this camera enabled direct visualisation of the passive tracer distribution within the porous plate. A blue colour-correction filter (HOYA 80B) is fitted to the lens to attenuate the longer-wavelength fluorescence emitted by the tracer on the porous plate, thereby increasing the relative contribution of the shorter-wavelength transmitted light reaching the camera sensor.

We found that continuous illumination, especially from blue light, caused significant photodegradation of the fluorescein over the duration of a typical washing experiment. Thus, to avoid photodegradation, illumination from the blue LEDs was restricted to only \SI{1}{\second} in each \SI{10}{\second} interval (or \SI{20}{\second} intervals for the longer experiments with the low permeability porous plate) during which an image was captured from the camera.  

Schematics of the whole experimental setup are shown in Fig.~\ref{fig:setup}(b). The washing solution was supplied to the washing apparatus by a gear pump (Cole-Parmer MasterFlex digital gear pump fitted with a P35 pump head) from a constant-head tank to prevent fluctuations in the incoming washing solution flow rate. The funnel at the end of the plate fed a second identical gear pump that drove the effluent through an in-line fluorometer (see Sec.~\ref{sec:fluorometer} for details) before discharging it into the effluent tank located on a high-capacity balance (Ohaus Defender 5000 XTREMEW). Both the fluorometer, which measured the fluorescein concentration in the effluent, and the \SI{150}{\kg} capacity balance (with a precision of \SI{0.01}{\kg}) used to determine the mass flow rate $Q$ in the film flow, operated with a sampling interval of \SI{2}{\s}.

Prior to the start of an experiment, we set the porous plate in a horizontal position and saturated it with the washing solution. A known mass of 0.1 w/w\% aqueous disodium fluorescein solution (containing mass $m_d$ of fluorescein) was placed on the surface. The fluorescein was allowed to dwell for a prescribed time period, $\tau$, on the porous plate, so that it could diffuse within the porous matrix. During the dwell period, a light-proof cover protected the fluorescein from photodegradation while preventing evaporation of water from the porous plate. The use of a light-proof cover improved significantly our results, as photodegradation of the fluorescein and the evaporation of water during the dwell period were found to have a substantial impact on the concentration measurements.

After the dwell period ended, we initiated the washing experiment by rapidly inclining the plate to an angle $\alpha$ and initiating the washing flow at $t=0$. To prevent the water film from breaking into rivulets on the glass region upstream of the porous surface, the flow emerging from the reservoir gate was first blocked on the glass boundary upstream and then guided with a squeegee by dragging it from the glass region onto the first centimetre at the top of the porous surface. At $t=0^-$ (just before inclining the plate), the surface manifestation of the tracer-rich region occupied an area $A_0$. Once established (for $t>0$), the washing fluid flowed first over the glass plate for approximately 2 cm upstream of the porous medium before flowing as a film with a thickness of approximately \SI{1}{\milli \metre} over the porous medium itself. The fluid then flowed over a second region of impermeable glass plate downstream before falling freely into a collecting funnel.

In all the experiments conducted in this study, the washing fluid mass flow rate was $Q=\SI{14.85\pm 0.26}{\gram \per \second}$. The mass of fluorescein that has been removed from the porous plate at time $t$ can be calculated as 
\begin{equation}
    m_r =  \int_{0}^{t} Q C(t') \, \dd t', 
\end{equation}
where $C(t)$ is the instantaneous concentration of the tracer in the effluent measured by the fluorometer. At the end of the experiments, we ensured that the effluent tank was well mixed and then recirculated some of the homogenised effluent through the fluorometer to verify mass conservation. In most cases, the mass of fluorescein determined by the fluorometer was slightly lower than the deposited mass, primarily due to unavoidable light-induced degradation of the fluorescent tracer from the imaging during washing. However, the tracer mass removed obtained from time integration of the fluorometer data was found to agree with the total tracer mass collected in the effluent tank within an average relative difference of approximately \SI{0.6}{\percent}. 

In the parametric study of this work, we varied the inclination angle $\alpha$ (6.5, 11.0, 14.8 and 20.4 degrees), the deposited tracer mass $m_d$ (0.2 or \SI{0.5}{\milli \gram}, corresponding to 0.2 or \SI{0.5}{\milli \litre} of tracer solution) and the dwell period $\tau$ (2 or \SI{18}{\hour}). We also varied the properties of the porous medium, such as permeability and surface roughness, by constructing porous plates with two different bead size distributions, 100-\SI{200}{\micro \metre} and 300-\SI{400}{\micro \metre}.

Preliminary tests showed that an important factor in ensuring experimental reproducibility was the degree of surface saturation of the porous plate before the deposition of the tracer. To maintain consistent initial conditions, the plate was not allowed to dry between experiments as drying would increase the chances of having different distributions of trapped air within the pore space. If the plate had dried, it was not used for experiments before running the washing film on it for approximately two hours and then letting it fully saturated with the washing solution for at least a day. Before each experiment, the plate surface was deliberately oversaturated by pouring washing fluid with a beaker, after which excess liquid was removed using a squeegee. This protocol provided great reproducibility.

The fluorescein solution was placed on the surface using a \SI{1}{\milli \litre} syringe by hand and dispensing around \SI{0.1}{\milli \litre} over a period of approximately \SI{2.5}{\second}. The solution spread rapidly over the wet porous surface in a circular pattern with radius of around \SI{30}{\milli \metre} and \SI{45}{\milli \metre} immediately after deposition for fluorescein solution volume of \SI{0.2}{\milli \litre} and \SI{0.5}{\milli \litre}, respectively. As noted above, the medium was then covered to reduce any evaporative loss and protect the tracer from photodegradation. This preparation protocol for the porous plate ensured  reproducibility between experiments. For each set of parameters, nominally identical experiments were repeated three to four times, with varying location of the tracer deposition to account slight inhomogeneities in the porous plate. Additionally, measurements showed that temperatures between 10 to \SI{50}{\celsius} do not affect concentration measurements from the fluorometer. However, experiments conducted with the washing fluid at \SI{17}{\celsius}, \SI{23}{\celsius}, and \SI{40}{\celsius} showed that temperature variations have a significant influence on tracer transport due to changes into the water dynamic viscosity and, potentially, tracer diffusivity with temperature. In this work, the washing fluid temperature was typically between \SI{20}{\celsius} and \SI{23}{\celsius}, to minimise the effect of temperature on mass transfer dynamics. This small variation in temperature is expected to lead to only a \SI{7}{\percent} variation in dynamic viscosity, and approximately 10 to \SI{12}{\percent} in tracer diffusivity \cite{Easteal1985}.

\subsection{\label{sec:plate}The porous substrate}

The substrate used in the experiments was purpose-made in-house by thermally sintering soda–lime glass beads within a mould formed by window glass plates. Prior to sintering, the beads were thoroughly washed to remove dust and then dried. The sintering was carried out in a glass fusing kiln (Nabertherm GF75), producing a mechanically stable composite porous plate. The final porous substrate had planar dimensions of $310 \times \SI{200}{\milli\metre\squared}$ and a thickness $\thick=\SI{6}{\milli\metre}$. The porous region was laterally bounded by solid glass plates of the same thickness and mounted on a flat glass base, as shown in Fig.~\ref{fig:setup}. The transitions between the porous region and the surrounding solid glass were nearly seamless, minimising any disturbance to the surface-washing film flowing above the plate. Further details of the fabrication procedure are provided in Appendix~\ref{sec:plate_fabrication}.

For this study, we fabricated plates of two different permeabilities, one using glass beads with diameters in the range 100--\SI{200}{\micro \metre} (supplied by Blast Spares Direct in 25 kg bags) and the other with diameters in the range 300--\SI{400}{\micro \metre} (supplied by Moleroda Finishing Systems in \SI{25}{\kilo \gram} bags). We henceforth refer to these media as `fine' and `coarse', respectively. The method we developed to measure the permeability $\permeability$ after sintering is presented in Appendix \ref{sec:permeability_measurement}. Specifically, the permeabilities of the plates were found to be $\permeability_f=(3.02 \pm 0.07) \times 10^{-12}$ and $\permeability_c=(1.64 \pm 0.49)\times 10^{-11}$ \SI{}{\metre \squared} for the fine and coarse media, respectively. 

The permeability ratio between the two plates, $\permeability_{c}/\permeability_{f}$, is $5.57 \pm 2.04$. For a homogeneous porous medium made of mono-disperse spherical beads with the same packing, the Kozeny--Carman equation suggests $\permeability \propto\dia_p^2$, where $\dia_p$ is the bead diameter. Using the mean diameters, $\bar\dia_f$ = \SI{150}{\micro \metre} for the fine and $\bar\dia_c$ = \SI{350}{\micro \metre} for the coarse beads, we would anticipate $\permeability_{c}/\permeability_{f}=5.44$. This value is similar to the ratio calculated from our permeability measurements, although the conditions for $\permeability \propto\dia_p^2$ do not exactly hold and some assumptions, explained in Appendix \ref{sec:permeability_measurement}, were made for the permeability measurement.

We elected to determine the porosity $\por$ of our porous substrate prior to sintering to eliminate potential issues of accurately measuring the volume of the sintered medium. Specifically, we filled a known volume, $V_{tot}$, with soda-lime glass beads and measured its mass, $m_{glass}$. Taking the density of soda-lime glass as $\rho_{glass} = 2.5$ \SI{}{\gram / \milli \litre}, the ratio of void volume over total volume, or porosity, $\por_{loose}$, can be calculated as $\por_{loose} = 1-(m_{glass}/\rho_{glass})/V_{tot}$. We measured $\por_{loose} =0.374 \pm 0.004$ and $0.368 \pm 0.007$ for the fine and coarse beads, respectively. The agreement in porosity measurement suggests that the bead shape and size distribution are similar in both groups of beads. However, after sintering, the porosity of the plate made from smaller beads might be lower due to the higher surface area of bead contacts, which leads to more fusion between smaller beads and a more compacted structure.

\subsection{Diagnostic methods}

\subsubsection{\label{sec:fluorometer} Measuring tracer concentrations in the effluent}

As noted in Sec \ref{sec:setup}, we performed accurate time-resolved quantification of the concentration of our tracer in the effluent from the washing flow for all the experiments. While for initial experiments we employed an absorbance spectroscopy approach (Uniqsis Flow-UV\texttrademark with a bespoke inline flow cell), this did not yield the necessary dynamic range. Additionally, measurements were highly sensitive to mechanical vibration of the flow cell and bubbles in the flow path, making it difficult to achieve consistent readings validated by mass conservation. Instead, we adapted a Hyperion Fluorescein Fluorometer (Teledyne Valeport) for our purposes. For fluorescein measurements the fluorometer had an excitation LED light of \SI{470}{\nano \metre} and a sensor to detect light of \SI{545}{\nano \metre}, emitted by fluorescein. The nominal dynamic range of the instrument was 0 - 500 ppb with a detectability limit of 0.03 ppb. 

\begin{figure*}[!ht]
\includegraphics[width=0.80\textwidth]{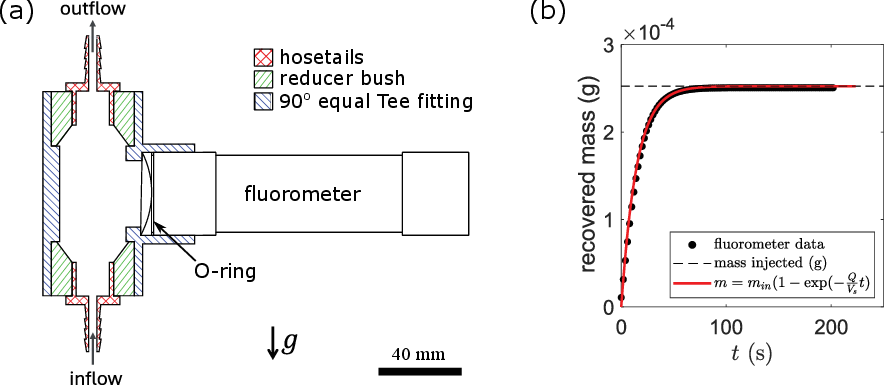}
\caption{\label{fig:fluorometer} (a) Drawing of the cross-section view of the Tee-connector with the end of the fluorometer fitted in it. (b) A test of the fluorometer response after the instantaneous injection of a known mass of fluorescein in the funnel.}
\end{figure*}

We installed this fluorometer, designed for oceanographic measurements on underwater moorings or vehicles, in the side branch of a standard \SI{40}{\milli \metre} PVC Tee-connector. The flow was directed along the length of the Tee-connector via reduction bushes at either end. A drawing of a cross-view of the Tee-connector with the end of the fluorometer fitted in it is shown in Fig.~\ref{fig:fluorometer}(a). The Tee-connector was oriented such that there was an upwards flow through it to prevent bubbles that could potentially affect the measurement from becoming trapped. Since the measurement conditions (primarily, the constrained measurement volume in the Tee-connector) differed from those for which the fluorometer was originally designed, we recalibrated the instrument by correlating the measured values with known fluorescein concentrations. We also observed that the orientation of the fluorometer about its axis in the Tee-connector had an influence on the measured concentration as its LED light source emitted at an angle rather than in line with its axis, causing the measurement volume (bounded by the Tee-connector) to vary with orientation. For our fixed fluorometer, orientation the fluorometer achieved a dynamic range of 0-1000 ppb with a sampling interval of \SI{2}{\second}. Additionally, one of the tests we undertook was to confirm that any fluorescein absorbed onto the PVC material of the Tee-connector components did not have a detectable effect on measurements. 

When using softened tap water for the aqueous fluorescein solution, we found that the calibration curves obtained on different days varied by up to approximately $\pm\SI{10}{\percent}$, as quantified by the deviation of each data point from the mean curve at each concentration. We identified day-to-day variations in water quality as the reason. It is well known that the water pH affects the spectroscopic properties of fluorescein \cite{Sjoback1995}. We therefore used a sodium carbonate/bicarbonate buffer (\SI{1}{\gram \per \kilo \gram} each of Na\textsubscript{2}CO\textsubscript{3} and NaHCO\textsubscript{3}) to regulate the water pH to around $9.87\pm 0.03$. This reduced the variation in the calibration curves to better than $\pm2\%$. If the raw (unsoftened) tap water was used then the solution would turn cloudy upon adding the buffer due to precipitation of calcium salts, which interferes with the fluorometer measurement. We confirmed that using reverse osmosis water (unbuffered) would give consistent fluorescein concentration measurements, but with the large volumes of washing fluid needed for the experiments this would be impractical. Additionally, the buffer lowered the surface tension of water, helping to maintain a more stable washing film that was less prone to breaking into rivulets and improving its penetration into the porous medium. 

To test the response of the fluorometer, we performed measurements in which a known mass of fluorescein, $m_{in}$, was injected almost instantaneously directly into the funnel. The volume of the system was $V_{s} = 220$ \SI{}{\milli \litre} with contributions from the funnel (\SI{120}{\milli \litre}), tubes (\SI{30}{\milli \litre}), and Tee-connector (\SI{70}{\milli \litre}). For assessing the performance, we approximated the system as a single well-mixed container with the instantaneous injection of fluorescein occurring at $t=0$, resulting in an initial concentration $C_0$. The inlet was supplied with fresh washing fluid with flow rate $Q = 15.15$ \SI{}{\gram \per \second} (here different than during the experiment) and the outlet with the same flow rate $Q$ had fluorescein concentration $C_{out}$. Assuming perfect mixing in the system at all times, we expect $C_{out} = C_{0}(\exp(-(Q/V_s)t))$. Therefore, the recovered cumulative mass from the fluorometer would be 

\begin{equation}
m=m_{in}(1-\exp(-(Q/V_s)t)).     
\label{eq:fluorometer_response}
\end{equation}
Figure~\ref{fig:fluorometer}(b) shows the cumulative recovered mass measured by the fluorometer, with this theoretical prediction superimposed. The excellent agreement between the theoretical curve and the experimental measurements, together with the nearly complete ($\simeq100\%$) mass recovery, confirms both the reliable performance of the fluorometer and the effective mixing within the system. Importantly, the measured response by the fluorometer is smeared in time due to the relatively large system volume. Note that the mass recovery was obtained also when the fluorescein was injected into the funnel using a syringe pump operating at a constant flow rate.

\subsubsection{Imaging} \label{sec:imaging}

The experimental setup for imaging our translucent media was designed to use a dye attenuation technique \cite{Cenedese1998,Allgayer2012}, similar to the approach in \cite{Landel2016} to quantitatively measure the depth-averaged tracer concentration field. When we are presenting images or results based directly on those, we correct for background variations in illumination using $I/I_0$, where $I$ is the light intensity field passing through the porous medium under consideration and $I_0$ is the corresponding intensity field when no dye is present. 
Our approach for quantification of the distribution of dye takes a form similar to the classical approach for dye attenuation. However, as detailed in appendix \ref{sec:camera_calibration}, the presence of the porous medium causes departures from the classical Beer--Lambert law. Nevertheless, except at the earliest times in the experiments, we are able to use
\begin{equation}
    \overline{C}\thick = -\scalefact\ln\left(\tilde{I}/\tilde{I}_{0}  \right),
    \label{eq:conc_BL}
\end{equation}
where $\overline{C}(x,y,t)$ is the depth-averaged concentration (here with dimensions of mass of solute per unit volume of fluid) and $h$ the thickness of the porous medium. Here, $\tilde{I}(x,y,t)$ and $\tilde{I}_0(x,y)$ are coarse-grained versions of $I$ and $I_0$, respectively, the coarse graining process removing the microstructure present in the raw images due to resolving individual particles.
Integrating $\overline{C}h$ over the area of the plate $A_p$ then gives the estimate for the mass of dye remaining in the plate
\begin{equation}
    m_{BL}(t) = \thick\int_{A_p} \overline{C}\,dx\,dy = -\scalefact\int_{A_p} \ln\left(\tilde{I}/\tilde{I}_{0}\right)\,dx\,dy,
    \label{eq:mass_BL}
\end{equation}
where the subscript $BL$ indicates that the mass is estimated from the Beer--Lambert law.

In contrast to the classical case, the scale factor $\scalefact$ (with dimensions of mass per unit area) is not simply a constant (see appendix \ref{sec:camera_calibration}) but depends on the particle size from which the medium is constructed and the dye concentration profile within the medium. However, by determining $\scalefact$ from a single-point calibration of a given set of control variables, we obtain close agreement with the fluorometer measurements (see Section \ref{sec:fluorometer}).

\section{\label{sec:results}Results and discussion}

\subsection{\label{sec:decontamination_flow}Transport of the passive tracer out of the porous plate}

\begin{figure*}[!ht]
\includegraphics[width=0.95\textwidth]{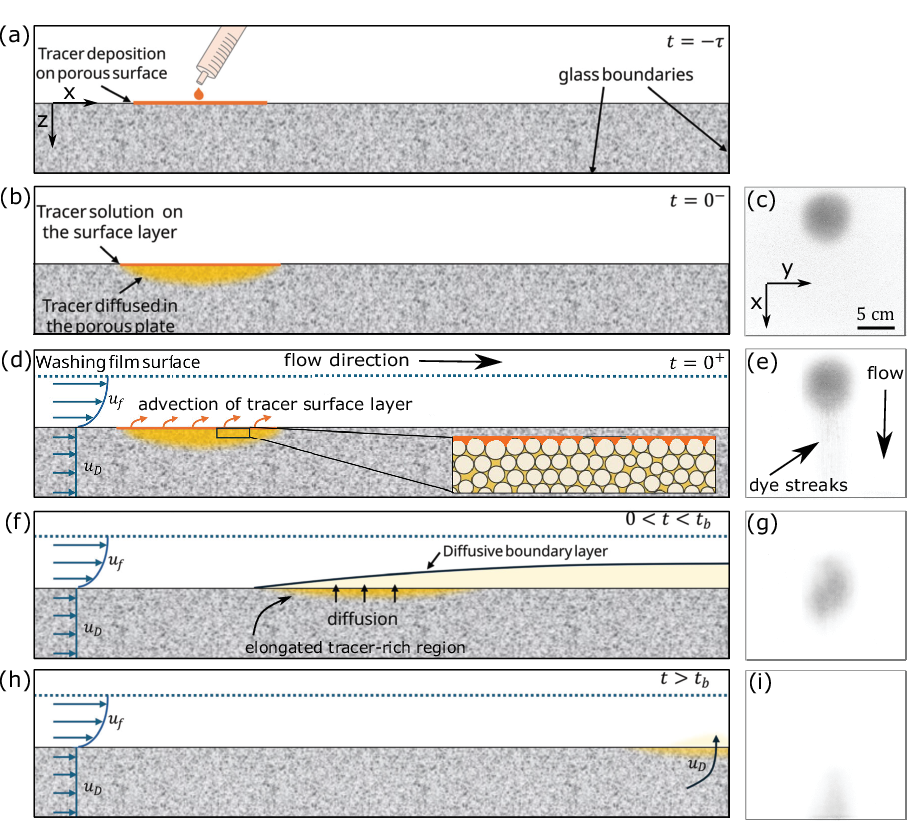}
\caption{\label{fig:washing_schematics}Schematics and corresponding experimental images of the stages of a passive tracer removal from the porous plate during surface washing. Left: Schematics showing cross-section view of the porous medium and the surface film flow illustrating how the distribution of tracer (in yellow and orange) evolves in time and space during the different stages of the washing process. Right: Representative background-subtracted experimental images of the porous plate at each stage obtained from direct visualisation using light absorption to reveal the tracer distribution \emph{in situ} the porous medium, during surface washing. Dark regions reveal higher concentration regions of the tracer within the porous plate. }
\end{figure*}

The removal of the passive tracer from the porous plate occurs in three stages in which different transport mechanisms are dominant. Figure~\ref{fig:washing_schematics} summarises these stages, showing schematics of the porous plate cross-section on the left and corresponding top-view (background-subtracted) experimental images of the tracer distribution (dark regions) during washing on the right. 

Fig.~\ref{fig:washing_schematics}(a) illustrates the deposition of the tracer solution onto the wet porous surface and its initial rapid spreading across it. During the subsequent dwell period, $\tau$, the tracer diffuses into the porous plate, forming a concentration distribution similar to that shown in Fig.~\ref{fig:washing_schematics}(b) and (c). At the surface of the porous plate, the tracer is found in the voids between the glass beads of the top layer indicated as dark orange region (see Fig.~\ref{fig:washing_schematics}(b)). The plan-view tracer distribution within the porous medium is roughly axisymmetric. In Fig.~\ref{fig:washing_schematics}(d) and (e), the tracer-rich surface layer comes in contact with the washing film almost at $t=\SI{0}{\second}$, after which it gets nearly instantaneously advected and carried downstream. We shall refer to this fast initial process as `surface flushing'. Dye streaks at the surface of the porous medium can be observed downstream of the deposition site (see Fig.~\ref{fig:washing_schematics}(e)). Once the top superficial layer of the tracer solution has been removed, the tracer that originally diffused deeper into the plate during $\tau$, begins to diffuse slowly back towards the surface and into the washing film, forming a tracer-rich diffusive boundary layer in the flow, as illustrated in Fig.~\ref{fig:washing_schematics}(f). Simultaneously, flow within the porous medium (Darcy-like flow \citep{Beavers1967}) induced by gravity advects the tracer patch downstream. As can be seen in Fig.~\ref{fig:washing_schematics}(g), streamwise elongation of the tracer patch occurs due to longitudinal hydrodynamic dispersion, as the tracer is transported through the porous medium. Deviations from a perfect ellipse are observed, probably due to local inhomogeneities in the porous medium. Finally, at time $t_b$, the tracer reaches the end of the porous plate, where it encounters the glass boundary and starts getting advected out of the porous plate by a surface-normal upward flow within it, as illustrated in Fig.~\ref{fig:washing_schematics}(h). In Fig.~\ref{fig:washing_schematics}(i), a long tail of tracer with reduced concentration can be seen upstream of the end of the porous plate. We expect that the flow near the downstream glass boundary of the porous plate becomes more complex and two-dimensional due to the no flow boundary condition, in contrast with the nearly uni-directional flow across the depth of the porous medium, upstream and sufficiently far from the end of the plate.

\begin{figure*}[!ht]
\includegraphics[width=1.0\textwidth]{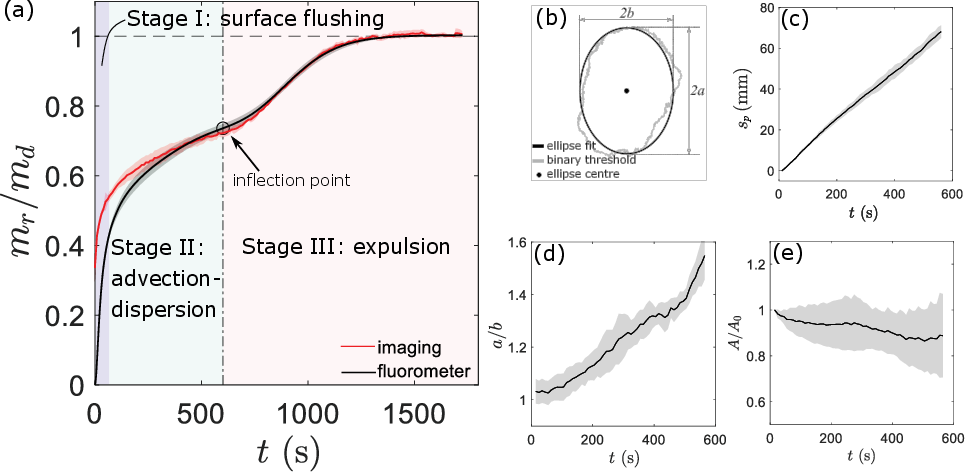}
\caption{\label{fig:decontamination_dynamics}
Graphs presenting the dynamics of passive tracer transport within and out of a porous plate (experiment E7 from Table~\ref{tab:parameters}). (a) The tracer mass removed from the porous plate, $m_r$, normalised with the tracer mass deposited on the porous plate, $m_d$, against time, $t$. Data from effluent concentration measurements by the fluorometer (black curve) and from dye attenuation (red curve). The three stages of the mass removal dynamics are shaded with different colours: (I) the surface flushing stage, (II) the advection--dispersion stage, (III) the expulsion stage. (b) A typical image of the tracer patch during Stage II, overlaid with the binary contour of the patch and the fitted ellipse. The major and minor axes of the ellipse are denoted by $a$ and $b$, respectively. (c) The distance, $s_p(t)$, travelled downstream by the centre of the ellipse fitted to the tracer patch in time. (d) The ratio of the major axis, $a$, over minor axis, $b$, of the ellipse fitted to the tracer patch plotted against time, $t$. (e) The normalised plan-view area of the tracer patch, $A/A_0$, against time, $t$. In all the plots the black solid lines represent the mean calculated from 3 experiments and the shade spans vertically 2 standard deviations.}
\end{figure*}

For a quantitative analysis of the transport dynamics, we examine a typical experiment in Fig.~\ref{fig:decontamination_dynamics} conducted with an inclination angle $\alpha=\SI{11.0}{\degree}$, the coarse porous plate, a plan-view tracer area of $A_0 = (2.66 \pm 0.69) \times 10^3\;\si{\mm\squared}$ at $t= \SI{0}{\second}$ ($m_d$ = \SI{0.2}{\milli \gram}), and a dwell period prior to washing of $\tau=\SI{18}{\hour}$ (experiment E7 from Table~\ref{tab:parameters}). The variation in $A_0$ here was determined from repeated experiments under identical conditions with the same tracer solution volume. Fig.~\ref{fig:decontamination_dynamics}(a) shows the total tracer mass that has been removed from the porous plate, $m_r$, normalised with the tracer mass initially deposited on the porous plate, $m_d$, as a function of time, $t$. 
We present curves obtained from both the fluorometer (black solid line) and dye attenuation (red solid line), with the imaging data scaled by a single multiplicative factor, $\scalefact$ (see Eq. \ref{eq:mass_BL}).
The indicated inflection point, which corresponds to the time $t_b$, was determined by locating the minimum of the first derivative of a least-squares spline fitted to the experimental data from the fluorometer. The three different stages of tracer removal after $t=0$, can be identified by the evolution of $m_r(t)/m_d$. The solid lines represents the mean curve obtained from three repeated experiments, and the shaded region spans two standard deviations. The mean and standard deviation of the removed mass were computed at each time instant. The use of shading to highlight the standard variation (based on the imaging measurements) between repeated experiments is also used in Fig.~\ref{fig:decontamination_dynamics}(c)--(e).

In Fig.~\ref{fig:decontamination_dynamics}(a) at early times, we observe fast removal of the tracer from the plate due to surface flushing (see Fig.~\ref{fig:washing_schematics}(d) and (e)). Although the flushing is nearly instantaneous, as it occurs in the advective time scale of the film ($L/u_f$, where $L$ is the length from the deposition site to the end of the porous plate, and $u_f$ is the film velocity), the flushed tracer remains in the volume of the fluorescence measuring system for a finite residence time before it is no longer detected (see Section \ref{sec:fluorometer}). Here, the system volume is $V_{s} = \SI{270}{\milli \litre}$ (funnel \SI{170}{\milli \litre}, tubes \SI{30}{\milli \litre}, and Tee-connector \SI{70}{\milli \litre}) and the inlet (washing solution) flow rate is $Q = \SI{14.85}{\gram \per \second}$, so for 95\% removal of the surface flushed tracer from the system, time $t_{flush} = t_{95\%} =\SI{54.5}{\second}$ is needed according to Eq.~\ref{eq:fluorometer_response}. In our porous plate experiments for $t \lesssim t_{flush}$, the measured concentration by the fluorometer is primarily due to tracer removed from the rapid initial flushing stage rather than  tracer removed in the slower second stage. Indeed, in Fig.~\ref{fig:decontamination_dynamics}(a) for $0\leq t\lesssim \SI{55}{\second}$, we observe that approximately 40\% of the original tracer mass has been removed. In contrast, the measurement from imaging (red line) responds more rapidly to the initial surface flushing because it directly records the tracer distribution on the porous plate.

After the top superficial layer of the tracer has been removed from the plate by surface flushing (Stage I) and until the indicated inflection point at time $t_b$, the tracer removal enters Stage II (Fig.~\ref{fig:washing_schematics}(c) and (g)). By comparing the evolution of the curve in Fig.~\ref{fig:decontamination_dynamics}(a) with image sequences showing the redistribution of the tracer within the porous plate (similar to the images showed in Fig.~\ref{fig:washing_schematics}(c),(e),(g),(i)), we observe that the point of inflection corresponds to the moment the leading edge of the tracer patch first reaches the downstream glass boundary of the plate.

Focusing on Stage II in Fig.~\ref{fig:decontamination_dynamics}(b)--(d) we examine the motion and shape of the plan-view of the tracer patch as it is transported downstream for times $t<t_b\approx \SI{600}{\second}$ (for this particular experiment). The boundary of the tracer patch is detected using a fixed threshold applied across all images during washing to ensure consistent segmentation/identification of the boundary. The patch boundary is then fitted with an ellipse, constrained to have its major axis oriented along the $x$-axis using a least-squares criterion. Figure~\ref{fig:decontamination_dynamics}(b) illustrates a tracer patch from a real image on which the boundary of the patch after image segmentation and the fitted ellipse are superimposed. The location of the centre of the ellipse along with its major and minor axis dimensions, $a$ and $b$, respectively, are also shown. Fig.~\ref{fig:decontamination_dynamics}(c) shows the downstream displacement of the centre of the patch (i.e. the centre of the ellipse) from the point of deposition until it approaches the downstream boundary of the porous plate at time $t_b$. Motion in the transverse $y$ direction perpendicular to the flow was found to be negligible. The patch displacement gives an average interstitial velocity of $u_p = \SI{0.121 \pm 0.001}{\milli \metre \per \second}$. As the tracer moves downstream, its shape changes, becoming elongated along the direction of the flow, as can bee seen in Fig.~\ref{fig:decontamination_dynamics}(d) where $a/b$ increases with time. The increase in $a/b$ as the tracer patch is advected downstream through the porous plate indicates substantial streamwise hydrodynamic dispersion. The streamwise dispersion is a consequence of the velocity differences along multiple flow pathways, which cause tracer particles to spread downstream \cite{Woods2014}. We have also studied the evolution of the patch plan-view area $A$ over time (the area is based on the segmented image not the fitted ellipse), hereafter referred to simply as the patch area. The patch area, $A$, generally decreases over time, but due to the large variations between experiments and the complex nature of this feature, the standard deviation is typically large. In addition to the dispersion-induced elongation that would cause area increase, diffusive removal through the interface into the washing film causes the patch to gradually fade and shrink.

During Stage II, tracer from deeper in the porous material is transported upwards and into the washing film by diffusion, which can be enhanced by transverse hydrodynamic dispersion in the surface-normal ($z$) direction depending on the P\'eclet number,
\begin{equation}
    Pe_p = \frac{u_p \ d}{D_{diff}},
\end{equation}
of the porous medium. Here, $u_p$ is the interstitial velocity, $d$ is a characteristic length (pore size or the grain size) and $D_{diff}$ is the molecular diffusion coefficient, which equals \SI{4.2e-10}{\m\squared\per\second} for disodium fluorescein diffusion in water \cite{Casalini2011}. While only longitudinal dispersion is readily measured from our experimental data in Stage II, surface-normal transverse dispersion may also be present and play a role in tracer removal from the porous plate. Transverse dispersion coefficient, $D_T$ is typically one to two orders of magnitude smaller than the longitudinal dispersion coefficient, $D_L$, and considerably more difficult to measure experimentally as data are typically highly scattered, but the two are related to the pore P\'eclet number $Pe_p$ in a similar way \cite{Woods2014,DeLigny1970}.

In a compilation of data from a broad number of studies, Delgado \cite{Delgado2007} reported that in unconsolidated porous media, the dispersion-to-diffusion coefficient ratio had a number of regimes depending on $Pe_p$: 
\begin{equation}
    \frac{D_T}{D_{diff}} \sim \left\{
    \begin{array}{cl}
        1, & \textrm{for }Pe_p<5, \\
        Pe_p^{1.1}, & \textrm{for }5 < Pe_p < 300, \\
        Pe_p, & \textrm{for }300 < Pe_p < 10^5.
    \end{array}
    \right.
    \label{eq:DT_ration_with_Pe}
\end{equation}
These empirical models for dispersion, developed from experimental observations are further supported by simulations \cite{Bijeljic2007}. For the present case of Fig.~\ref{fig:decontamination_dynamics}, we estimate the P\'eclet number as $Pe_p \approx 100$ using a superficial velocity $u_p = \SI{0.121}{\milli \metre \per \second}$, a characteristic length scale given by the mean glass bead diameter, $\bar\dia_c =$ \SI{350}{\micro \meter}, and a molecular diffusivity $D_{diff}=\SI{4.2e-10}{\m\squared\per\second}$. Although the porous medium here is consolidated, and the direct applicability of these empirical correlations may therefore be uncertain, the estimated P\'eclet number is sufficiently large to consider that vertical mass transfer is dominated by dispersion.

For $D_T \gg D_{diff}$, tracer removal occurs through the combined action of advection and hydrodynamic transverse dispersion. While molecular diffusion enables tracer exchange between neighbouring pore-scale streamlines, in the case of high $Pe_p$ velocity variations within the porous structure amplify this effect, producing an effective transverse dispersive flux towards the surface. This process is driven by the strong concentration gradient between the tracer-rich porous medium and the clean washing film and by the heterogeneity of pore-scale flow pathways, which causes tracer particles to follow different trajectories and spread normal to the mean flow direction \cite{Woods2014, Whitaker1999}. Hence, we shall refer to Stage II as the `advection--dispersion' stage.

Returning to Fig.~\ref{fig:decontamination_dynamics}(a), Stage III begins at $t=t_b$ and the dominant tracer removal mechanism becomes advection within the porous medium driven by the vertical component of the two-dimensional flow field generated by edge effects at the downstream boundary of the plate (see Fig.~\ref{fig:washing_schematics}(h) and (i)). Owing to the change in removal mechanism to advection, the slope in Fig.~\ref{fig:decontamination_dynamics}(a) increases rapidly, before it gradually decreases, following a second inflection point (not indicated on the figure), as the less concentrated tail of the tracer patch is removed from the plate. We shall refer to Stage III as the `expulsion' stage.

At the end of the washing process, the fluorometer measurements of the mass recovered form the porous plate, $m_r$ is typically within $\pm \SI{1.5}{\percent}$ of the deposited mass $m_d$ for all our experiments. As this number is relatively low and comparable with the uncertainty of the calibration curve, this confirms that all the tracer mass initially present in the porous medium is being removed completely during the washing process, while it highlights the high accuracy of the fluorometer effluent analysis. 

\subsection{\label{sec:parametric_study}Impact of key parameters on mass transfer dynamics}

We examine the effect of different parameters on the mass transfer dynamics: the inclination angle of the plate, $\alpha$, the dwell time period, $\tau$, the amount of tracer, $m_d$, (which is closely related to the initial surface area of the tracer distribution, $A_0$), and the permeability of the porous plate, $\permeability$. The experiments performed in this study and their parameters are summarised in Table~\ref{tab:parameters}.

\setlength{\tabcolsep}{8pt}
\begin{table}[h]
\centering
\caption{Experimental parameter combinations. Cells with check mark indicate the conditions of each experiment. Experiments under the same conditions were repeated 3 to 4 times.}
\setlength{\arrayrulewidth}{0.8pt}
\renewcommand{\arraystretch}{1.2}
\label{tab:parameters}

\begin{tabular}{|c|c|c|c|c|c|c|c|c|c|c||c|}

\hline
&\multicolumn{10}{c||}{Control parameters} & \multicolumn{1}{c|}{Output} \\
\hline

&\multicolumn{4}{c|}{$\alpha$ (degrees)} & \multicolumn{2}{c|}{$m_d$ (\SI{}{\milli \gram})} & \multicolumn{2}{c|}{$\tau$ (\SI{}{\hour})} & \multicolumn{2}{c||}{$\permeability \times 10^{12}$ (\SI{}{\metre \squared})} & $Pe_p$ \\
\cline{2-11}
Exp. & 6.5 & 11.0 & 14.8 & 20.4 & 0.2 & 0.5 & 2 & 18 & 3.02  & 16.4 &  \\
\hline
E1 & \checkmark &  &  &  &  & \checkmark & \checkmark & &  & \checkmark  & 88 \\
\hline
E2 &  & \checkmark &  &  &  & \checkmark & \checkmark &  &  & \checkmark   & 127 \\
\hline
E3 &  &  & \checkmark &  &  & \checkmark & \checkmark &  &  & \checkmark   & 163 \\
\hline
E4 &  &  &  & \checkmark &  & \checkmark & \checkmark & &  & \checkmark   & 242 \\
\hline
E5 &  & \checkmark &  &  &  & \checkmark &  & \checkmark &  & \checkmark   & 101 \\
\hline
E6 &  &  &  & \checkmark &  & \checkmark &  & \checkmark &  & \checkmark   & 227\\
\hline
E7 &  & \checkmark &  &  & \checkmark &  &  & \checkmark &  & \checkmark   & 101 \\
\hline
E8 &  & \checkmark &  &  & \checkmark &  &  & \checkmark & \checkmark  &  & 12 \\
\hline
E9 &  & \checkmark &  &  &  & \checkmark &  & \checkmark & \checkmark  &   & 12 \\
\hline
\end{tabular}
\end{table}

\begin{figure*}[!ht]
\includegraphics[width=0.92\textwidth]{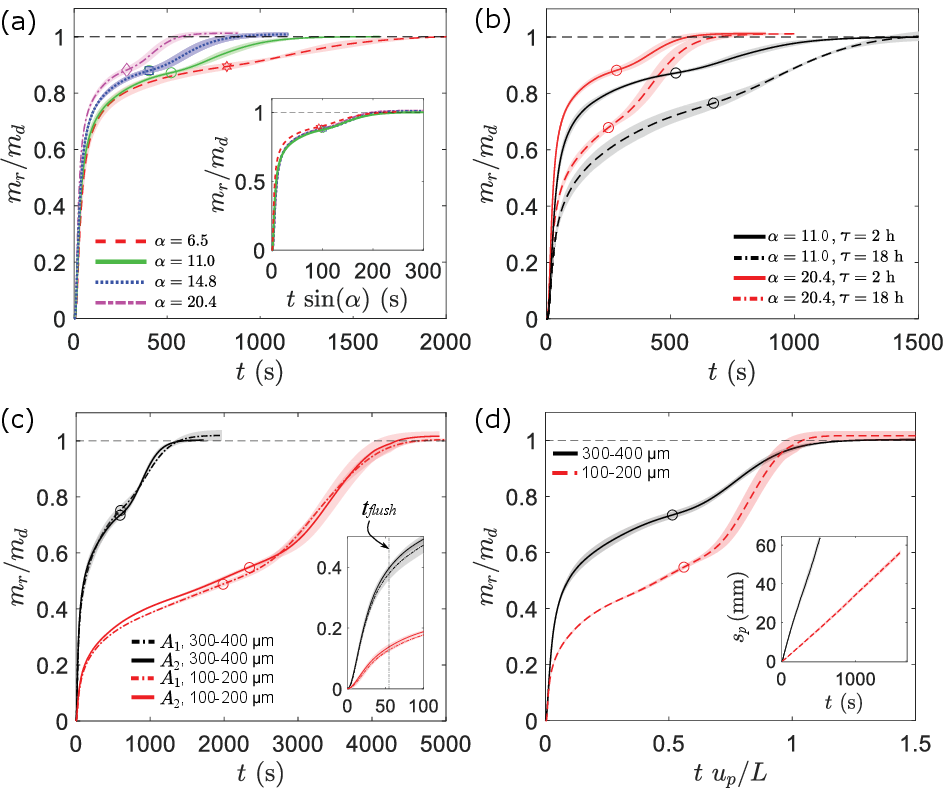}
\caption{\label{fig:parametric_analysis} The normalised mass of the tracer removed from the porous plate against time for varying parameters. (a) Varying inclination angles, $\alpha$ ($\tau$ = 2 h, coarse porous plate, $m_d$ = 0.5 mg). Experiments E1, E2, E3, and E4. The symbols indicate the inflection point, at time $t_b$. Inset: Rescaled time axis with $\sin(\alpha)$. (b) Varying $\tau$, 2 and 18 h, for two different inclination angles, $\alpha$ (coarse porous plate, $m_d$ = 0.5 mg). Experiments E2, E4, E5, and E6. (c) Varying area of initial contamination, $A_0$, $A_1 = (5.54 \pm 0.48) \times 10^3$\;\si{\mm\squared} and $A_2 = (3.07 \pm 0.51)  \times 10^3$\SI{}{\milli \metre \squared}, for two plates with different permeability (coarse and fine plates), ($\tau$ = \SI{18}{\hour}, $\alpha$ = \SI{11.0}{\degree}). Experiments E5, E7, E8, and E9. Inset: Close-up at short times. (d) Varying porous plate permeability using the coarse (300--\SI{400}{\micro \meter} beads) and fine (100--\SI{200}{\micro \meter} beads) plates ($\tau$ = \SI{18}{\hour}, $\alpha =\SI{11.0}{\degree}$, $m_d$ = \SI{0.2}{\milli \gram}). Experiments E7, and E8. Time rescaled with $u_p/L$. Inset: Patch travelled distance against time on two different plates. In all the plots the lines represent the mean from at least 3 experiments and the shade spans vertically 2 standard deviations.}
\end{figure*}

The results of the parametric study are presented in Fig.~\ref{fig:parametric_analysis}, where the variation of the chosen parameters make the primary removal mechanisms in each stage more apparent. Fig.~\ref{fig:parametric_analysis}(a) shows the relative tracer mass removed from the plate, $m_r/m_d$, as a function of time for four different inclination angles ($\alpha=$ 6.5, 11.0, 14.8 and \SI{20.4}{\degree}) while keeping the remainder of the experimental conditions the same ($\tau$ = \SI{2}{\hour}, coarse porous plate, $A_0=$ \SI{4.24\pm 0.83e3}{\milli \metre \squared} ($m_d$ = 0.5 mg)). Here, for simplicity, we have presented only the fluorometer measurements but the dye absorption measurements show a similar behaviour when a following a single-point calibration. Since the flow rate $Q$ does not change in our experiments, a steeper inclination angle causes the washing film to be faster and thinner. Steeper inclination angles also result in greater flow velocities within the porous medium and thus faster transport of the tracer patch downstream resulting in smaller $t_b$, as can be seen in Fig.~\ref{fig:parametric_analysis}(a). Similarly, for $t>t_b$, the slope of the curve is steeper for greater $\alpha$, since mass removal in the expulsion stage (Stage III) is advection-dominated. We can assess the impact of advection within the porous medium by rescaling time with $\sin(\alpha)$, to account for the down-slope component of gravity. The re-scaled results are shown in the inset in Fig.~\ref{fig:parametric_analysis}(a), where we see the four curves collapse in all three stages of the removal process. The overall shape of the curves and the location of the inflection points are within the standard deviation (plotted with shades of the same colour). Remarkably, regardless of the value of $\alpha$, the relative tracer mass removed from the porous plate until time $t_b$ is around 88\%, despite the advection--dispersion stage (Stage II) having a shorter duration for steeper $\alpha$.

While the scaling with advective velocity is consistent with expectations for the Stage III advection-dominated mass removal, the collapse of the curves (shown in the inset of Fig.~\ref{fig:parametric_analysis}(a)) during the advection--dispersion stage (Stage II) is noteworthy. This may be partly attributed to the faster surface-washing film, which sustains a larger concentration gradient across the interface between the porous plate and the washing film, thereby enhancing mass transfer. However, diffusive transport is independent of the angle, and so, for longer times $t_b$, would yield a larger $m_r$, which is not what we find.

Instead, the collapse of the curves in the inset of Fig.~\ref{fig:parametric_analysis}(a) suggest that the mass removal rate during Stage II is linked to the pore-scale velocity in a linear way. This is further evidence that geometric dispersion dominates over molecular diffusion. One can see this by assuming that tracer movement within the pore space is described by the advection--diffusion--dispersion equation,
\begin{equation}
    \frac{\partial C}{\partial t} + u_p\frac{\partial C}{\partial x} = (D_{diff}+D_L)\left(\frac{\partial ^2 C}{\partial x^2}\right)+(D_{diff}+D_T)\left(\frac{\partial^2 C}{\partial z^2}\right),
\end{equation}
where $x$ is the stream-wise direction, $z$ the surface-normal direction and $D_T$ and $D_L$ the transverse and longitudinal dispersion coefficients, respectively. Since removal from the porous medium can occur in the surface-normal direction only, the removal rate scales as
\begin{equation}
    \frac{C}{t} \sim (D_{diff}+D_T)\frac{C}{z^2}.
\end{equation}
Since we found $5 < Pe_p < 300$ in our experiments (see Table~\ref{tab:parameters}), $D_T/D_{diff}\sim Pe_p^{1.1}$ (see Eq.~(\ref{eq:DT_ration_with_Pe}). Therefore, we expect for the removal rate
\begin{equation}
    \frac{C}{t}\sim u_p^{1.1},
\end{equation}
i.e. to be almost linear.

Consequently, a surface-washing film with the same velocity $u_f$ but negligible Darcy velocity, i.e. $Pe_p\ll 1$, would be significantly less effective, as tracer removal would then rely primarily on diffusion rather than transverse dispersion.

Fig.~\ref{fig:parametric_analysis}(b) examines two different dwell periods, $\tau=\SI{2}{\hour}$ (solid lines) and $\tau=\SI{18}{\hour}$ (dashed-dotted lines), for two different inclination angles, $\alpha = \SI{11.0}{\degree}$ (black lines) and \SI{20.4}{\degree} (red lines). The remaining parameters are identical between these two experiments (coarse porous plate, $m_d$ = \SI{0.5}{\milli \gram}). This difference in dwell period is expected to lead to a difference in the sodium fluorescein diffusion distance, $L_{diff}\sim \sqrt{D_{e}\tau}$, where $D_e$ is the effective diffusivity in the porous plate of disodium fluorescein in water. The effective diffusivity is related to the molecular diffusivity by $D_e = D_{diff} \por/T$ with tortuosity $T$ typically around 1.4 for 3D packing of spheres  \citep{Gunn1987, Holzer2023}. Taking $D_{diff} \approx \SI{4.2e-10}{\m\squared\per\second}$ \cite{Casalini2011} and $\por \approx 0.37$, we obtain $D_e\approx \SI{6.3e-11}{\m\squared\per\second}$. We calculate that the difference in dwell period (2 or 18 hours) lead to a \SI{4.3}{\milli \metre} difference in the fluorescein diffusion distance. As greater $\tau$ means the tracer is diffused at greater depths into the porous plate, consequently, the concentration of the tracer adjacent to the surface will also be lower. Due to the lower concentration of tracer at the surface, the initial fast surface flushing (removal of the tracer surface layer) removes approximately 34\% less mass for $\tau=\SI{18}{\hour}$ than for $\tau=\SI{2}{h}$ for both inclination angles of Fig.~\ref{fig:parametric_analysis}(b). 

Since larger $\tau$ leads to less tracer mass being removed during the initial surface flushing (Stage I), more mass remains available for subsequent stages. Although at early times in the advection–dispersion stage (Stage II) the tracer concentration gradients within the porous plate are larger for shorter dwell periods, this trend reverses as the residual tracer mass decreases. At later times, longer dwell periods produce larger characteristic concentration gradients across the thickness of the porous plate. Because the removal rate is primarily controlled by dispersion, which increases with this gradient, the slope becomes steeper for $\tau=\SI{18}{\hour}$ than for $\tau=\SI{2}{\hour}$ as $t_b$ is approached at the same interstitial velocity (i.e., identical $\alpha$), as seen by comparing lines of the same colour in Fig.~\ref{fig:parametric_analysis}(b).

During the dwell period, the tracer diffuses farther also in the planar directions $x$ and $y$ for longer $\tau$. Specifically, we measured that $A_0 = (4.2 \pm 0.8) \times 10^3$ and $(5.4 \pm 0.7) \times 10^3\;\si{\mm\squared}$ for $\tau=\SI{2}{\hour}$ and \SI{18}{\hour}, respectively. This yields a $4.90 \pm 4.4$ mm difference in their radii, agreeing well with our theoretical estimations for the one-dimensional diffusion length difference between the two different dwell periods. Nevertheless, this difference in the initial extent of the patch does not affect significantly the arrival time at the downstream boundary, $t_b$, as shown by comparing the inflection point of the solid and dashed curves in Fig.~\ref{fig:parametric_analysis}(b).

The duration of the expulsion stage (Stage III) is independent of the tracer mass remaining in the porous plate at $t_b$. Because removal in the expulsion stage is governed by Darcy-like advection within the porous plate, its duration may vary with parameters such as the inclination angle $\alpha$ or the permeability (see, for example, lines of the same colour in Fig.~\ref{fig:parametric_analysis}b). Consequently, provided the flow conditions within the plate remain unchanged, the duration of this stage remains approximately constant.

In Fig.~\ref{fig:parametric_analysis}(c), we examine whether a larger tracer patch can lead to faster removal of the tracer mass from the plate into the washing film, since we may expect the mass transfer rate to be proportional to the patch area. We performed sets of experiments depositing two different volumes, \SI{0.5}{\milli \litre} ($m_d= 0.5$ mg) and \SI{0.2}{\milli \litre} ($m_d= 0.2$ mg) of the fluorescein solution on the porous plate, which led to fluorescein patches of two different sizes with area $A_1$ and $A_2$. Averaging the area measurements at $t=0$ following \SI{18}{\hour} dwell times from multiple experiments in both the coarse and fine plates, we calculate $A_1 = \SI{5.54 \pm 0.48 e3}{\milli \metre \squared}$ and $A_2 = \SI{3.07 \pm 0.51 e3}{\milli \metre \squared}$ for 0.5 and \SI{0.2}{\milli \litre} of the fluorescein solution, respectively. The spatial extent of the tracer distribution during the dwell period remains similar for both plates (coarse and fine) when the same tracer volume is applied. This implies that statistically our plates have similar structure. In some cases, especially in anisotropic porous media, lower permeability is associated with higher tortuosity \cite{LI2019}. An increase in tortuosity, which measures how much longer the actual transport path of the fluid or solute is compared with the straight-line (Euclidean) distance between two points, reduces the effective diffusivity within the porous network and would therefore delay tracer spreading from the point of deposition. However, here there is no evidence that such an effect occurs as differences in permeability do not significantly influence diffusive transport in the present system during the dwell period. 

The experiments of Fig.~\ref{fig:parametric_analysis}(c) show that, for a given porous plate, the removal rate is not significantly affected by the initial tracer volume or spreading area, as the solid and dot–dashed curves collapse for the same permeability (same colour). This indicates that mass transfer during Stages I and II is effectively one-dimensional and predominantly surface-normal, with the plan area playing a negligible role. In the present case, where $u_D \propto \sin(\alpha)$ is the same across all cases shown, tracer removal is primarily controlled by down-slope advection within the porous plate (through transverse dispersion in Stage II and expulsion in Stage III) rather than by molecular diffusion. Therefore, different initial patch areas do not influence the dimensionless removal rates, and the two cases with $A_1$ and $A_2$ exhibit indistinguishable mass transfer dynamics.

A further observation that can be seen in Fig.~\ref{fig:parametric_analysis}(c) is that experiments using plates with different permeability while keeping the other parameters identical reveal significant differences in all three stages of washing. During surface flushing (Stage I), in the coarse plate (black curves), the initial surface flushing removes approximately twice the mass of tracer compared to the fine plate (red curves). Although the porosity between the two plates is not expected to differ significantly (as discussed in Sec.~\ref{sec:plate}), the available void volume within the surface roughness layer (orange part in the zoomed region of Fig/~\ref{fig:washing_schematics}(d)), which can be swept by the washing film at $t=\SI{0}{\second}$ is greater for larger beads. Specifically, to estimate this void volume, we consider the upper half of the surface layer of a close-packed arrangement of uniformly sized spheres. For each spherical bead of diameter $\dia_p$ the free volume at the surface is $\frac{1}{2}(1 - \pi/6) \dia_p^3$ and the planar area it occupies, assuming a close packing, is $\dia_p^2$. The surface void volume, $V_{surf}$, per unit area is then
\begin{equation}
V_{surf} = 
 \frac{1}{2}\left(1-\frac{\pi}{6}\right),
 \label{eq:surface_void}
\end{equation}
simply scaling with the particle diameter. If we assume $\dia_p = \SI{350}{\micro \metre}$ and $\SI{150}{\micro \metre}$ for the higher (coarse beads) and lower (fine beads) permeability plates of Fig.~\ref{fig:parametric_analysis}(c), respectively, the ratio of the surface void volume of bigger to smaller beads is calculated to be 2.3. This can explain the difference in mass removal between the plates made of different bead size during surface flushing. In the inset of Fig.~\ref{fig:parametric_analysis}(c) the ratio of the removed mass from larger- over smaller-bead plate is $\sim 2.9$, not far from the value of 2.3 estimated above.

Tracer removal from the lower permeability plate (red curves in Fig.~\ref{fig:parametric_analysis}(c)) takes significantly longer to complete than from the higher permeability plate (black curves). However, when the time is rescaled by the interstitial velocity over the distance from the deposition site to the end of the porous plate $u_p/L$ ($L=\SI{14}{\centi \meter}$) as shown in Fig.~\ref{fig:parametric_analysis}(d), both the inflection point and the overall (rescaled) cleaning duration align for the two cases (we show only the cases with $A_2$ for clarity). Also, the removal rate in the advection--dispersion stage (Stage II) scales with $u_p/L$, as indicated by the nearly parallel curves before $t_b$. This scaling indicates that tracer removal is enhanced by surface-normal transverse dispersion and depends approximately linearly on $u_p$. In contrast, diffusion is not influenced by $u_p$ and as such we would expect these curves to be essentially parallel without rescaling by $u_p$. The similar rescaled removal rates during Stage II for plates of different permeability also suggest structural similarity at the microscale (degree of heterogeneity, pore connectivity, and surface-film accessibility), highlighting the reproducibility of our fabrication protocol.

As explained previously, the interstitial velocity, $u_p$, for each plate was determined by tracking (in time) the centre of the ellipse fitting the patch, $s_p(t)$, as shown in the inset of Fig.~\ref{fig:parametric_analysis}(d). As our media are isotropic, the interstitial velocity is proportional to Darcy velocity via porosity $\por$. Since the Darcy velocity is proportional to permeability, we expect $u_p \propto \permeability$. Indeed, the velocity ratio $u_{p_{c}}/u_{p_{f}} = \SI{0.1207 \pm 0.0007}{\milli \metre \per \second}/\SI{0.03219 \pm 0.0001}{\milli \metre \per \second} = 3.75 \pm 0.03$, agrees reasonably well with the permeability ratio $\permeability_{c}/\permeability_{f} =5.57 \pm 2.04$. 

Further image-based measurements of the removed tracer as a function of time can be found in Appendix~\ref{sec:dye_attenuation_results}, while additional image-based measurements of the tracer patch area and interstitial velocity for different inclination angles and permeabilities, which further highlight the dominant role of subsurface advection, are presented in Appendix~\ref{sec:parametric_measurements}.

\section{\label{sec:conslusions}Discussion and Conclusions}

In this work, we investigated the mass transfer dynamics of a passive tracer removed from a water-saturated porous plate subjected to surface washing, focusing on the case in which the pore fluid and the washing fluid are identical and fully miscible with the tracer. Although this configuration may appear simple, it represents a fundamental scenario for understanding transport mechanisms during the removal of a passive tracer from a porous medium by a surface washing flow, and provides a controlled experimental analogue for a wide range of practical cleaning and decontamination processes. Through a combination of qualitative and quantitative diagnostics, we achieved detailed temporal and spatial tracking of the tracer both within the porous medium and in the effluent. The experiments allowed us to quantify the tracer removal rate under a range of conditions and, through direct imaging, to visualise how the washing flow advects, disperses, and ultimately removes the tracer from the porous substrate.

The results show that mass transfer of a passive tracer miscible with the washing fluid during the washing process occurs in three distinct stages: an initial rapid mass removal from the plate due to surface flushing (Stage I), followed by the slower advection--dispersion stage (Stage II), during which the tracer is removed primarily by surface-normal transverse dispersion, and, finally, an acceleration of removal due to advection-driven transport within the porous plate during the expulsion stage (Stage III) when the tracer-rich patch reaches the downstream boundary of the plate. 

A parametric study revealed the influence of key factors on the mass transfer dynamics. Steeper inclination angles, $\alpha$, were shown to lead to more efficient tracer removal, both in terms of cleaning time and washing fluid usage. The observation that cases with different $\alpha$ exhibit self-similar mass transfer behaviour when time is scaled with the downstream component of gravity demonstrates the dominant role of advection within the porous medium in advection--dispersion and expulsion stages. In particular, during the advection--dispersion stage (Stage II), advective flow within the porous medium enhances tracer transport toward the surface via transverse dispersion, thereby accelerating the overall removal process beyond what diffusion alone could achieve.

Additionally, the dwell time, which determines how deeply the tracer diffuses into the porous plate prior to washing, plays a key role in the removal dynamics of each stage, as revealed by the parametric study. Longer dwell times lead to lower surface concentrations and longer surface-normal transport distances and therefore reduce the mass removed during the surface flushing (Stage I). The removal rate in the advection--dispersion stage (Stage II) depends on the amount of tracer remaining in the plate, with higher remaining mass resulting in higher removal rates. The final expulsion stage (Stage III) is also controlled by the interstitial flow, and the the removal rate, as in the advection--dispersion stage, increases with the amount of tracer remaining in the plate, while its duration is independent of the remaining tracer mass.

When deposited in a compact manner, the removal rate is insensitive to the initial tracer volume or spreading area, highlighting that Stages I and II are dominated by one-dimensional, surface-normal, advection-driven transport rather than surface-area effects.

Moreover, this study demonstrated that porous plates with lower permeability require longer times for complete removal of the tracer compared to those with higher permeability. This is mainly due to the lower interstitial velocity that makes removal by advection--dispersion (Stage II) and expulsion (Stage III) slower. Scaling the time with the interstitial velocity led to similar permeability-independent removal rates during Stage II, confirming that removal during this stage is enhanced by transverse dispersion, as diffusion alone would be insensitive to interstitial velocity. The plates also differed in surface roughness, with the lower-permeability plate exhibiting a smoother surface, due to the smaller scale of the surface features. This reduced the amount of tracer residing in surface voids and consequently decreased the mass removed during surface flushing (Stage I).

The closely similar removal rates during the advection-dispersion stage for different permeability plates when time is scaled with the interstitial velocity demonstrate the reproducibility of our porous plate fabrication protocol, confirming that the plates with different permeability have statistically similar structural characteristics, including pore connectivity, degree of heterogeneity, and surface accessibility. Moreover, one important benefit of our fabrication approach is that it offers considerable flexibility, allowing us to systematically vary parameters such as permeability, introduce layered structures with different pore sizes, or produce plates of tailored dimensions. This adaptive methodology provides the capability for future studies to explore how complex porous media influence mass transfer dynamics.

Our experiments highlight the complexity of practical situations where boundaries and heterogeneities in material composition or properties are present. Indeed, all real-world porous structures have finite dimensions with boundaries that can trigger a transition in the dominant mass-transfer mechanism, from Stage-II–type advection–-dispersion to Stage-III–type advection--dominated removal, thereby changing the removal rate over the course of the cleaning process. Many materials, such as concrete, have regions of different porosity and pore size (often exhibiting a hierarchy of pore scales) \cite{Kumar2003}, with pore size ranging from roughly the nanometre to the millimetre scales. In such regions, the P\'eclet number decreases, making the much slower molecular diffusion the primary mechanism during Stage II and eliminating dispersion enhancement. Furthermore, Stage III might not happen at all as at very low pore P\'eclet number, all the mass will be removed from the porous medium by diffusion before the patch reaches the downstream boundary of the porous domain.

Beyond the experimental observations, this study also introduces a novel, high-precision, high-accuracy, low-cost method for real-time measurement of disodium fluorescein concentration in the continuous flow of the effluent (see Sec. \ref{sec:fluorometer}) that provides a practical tool for future investigations of similar systems. Additionally, we dealt with challenges related to direct in-situ visualisation of porous media flows. While advanced imaging techniques such as magnetic resonance imaging (MRI) \cite{Sederman2001,Li2009,Barrie2000} or X-ray computerised tomography (CT) \cite{Zhan2000} can provide detailed structural and flow information, they are limited by small test volumes and slow acquisition times, while being very expensive and specialized tools. In this work, direct top-down imaging of light absorbance from the tracer provided macroscopic visualisation of the transport and revealed key dynamics, such as the elongation of the tracer patch in the flow direction as it is advected downstream, a behaviour consistent with longitudinal dispersion. Although not fully quantitative, our method is simple, cheap, well resolved in time, and yielded valuable semi-quantitative insights that complemented the mass removal measurements from a fluorometer, demonstrating its suitability for capturing the key features of surface washing in porous media. Overall, the high temporal resolution and reliability of our diagnostic methods, which enable detailed tracking of tracer transport and removal, provide a valuable methodology for experimental studies in porous media.

Our findings enhance the understanding of mass transfer in porous materials under the influence of an external flow and provide guidance for improving surface-washing strategies in industrial and environmental cleaning and remediation contexts. The combination of high-resolution diagnostics and controlled experiments allowed us to isolate the fundamental transport mechanisms limiting tracer removal. This approach provides clarity on processes that are often difficult to access experimentally and tackles some of the technical and scientific challenges highlighted in \cite{Landel2021,Wilson2022}. Importantly, it establishes a quantitative foundation that can inform the studies of more complex scenarios of porous media surface washing, involving initially unsaturated porous media, partial miscibility, wettability effects, non-uniform initial conditions, or reactive processes, thereby supporting the development of more effective strategies for cleaning and decontaminating real-world porous materials.

\begin{acknowledgments}
The authors acknowledge funding from the UK's Defence Science and Technology Laboratory (Dstl) through contracts DSTL0000030131 and DSTLX1000138254. The authors also thank the late Henry McEvoy for his contributions to the conceptualisation of this project and for his efforts in securing the financial support that made it possible. The authors are grateful to Dr Rich Thomas and Dr Steve Marriott for their contributions to the development of the buffered system for accurate fluorescein concentration measurements. 

\end{acknowledgments}

\appendix

\section{Porous plate fabrication} \label{sec:plate_fabrication}

A glass mould is created by gluing pieces of plate glass onto a plate glass base using Bullseye GlasTac Glue (an adhesive based on hydroxypropyl methylcellulose that burns away at high temperatures). Once the glue is fully cured, the mould is filled with glass beads and the excess is scraped with a rod to ensure a flat surface before placement in a glass fusing kiln (Nabertherm GF75). In the glass kiln, the plate glass pieces and glass beads fuse or sinter together to create a porous matrix. The porous matrix, together with its mould, is used in the experimental setup as a whole (see Fig.~\ref{fig:setup}).

Preparation for porous plate fabrication includes careful cleaning to remove dust and oils from the plate glass, and pre-washing of the soda-lime glass beads with detergent and reverse osmosis water as well as their thorough rinsing in more reverse osmosis water and drying in a laboratory oven. The washing of the glass beads removes dust and other impurities, which could otherwise cause uneven heat distribution, inconsistent fusion between the beads, reduced mechanical strength, and non-uniform porosity with larger gaps between beads.

\begin{figure*}[!ht]
\includegraphics[width=0.95\textwidth]{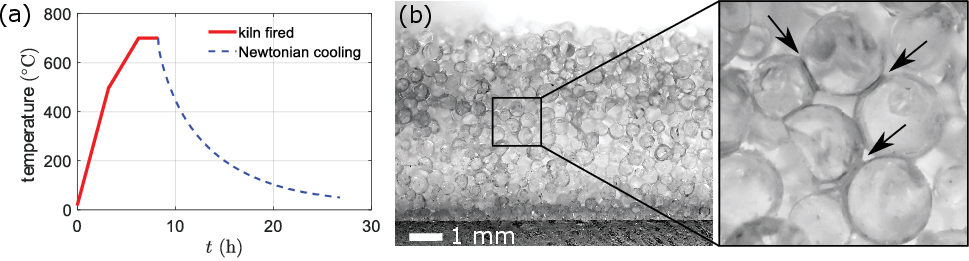}
\caption{\label{fig:kiln_thermal_cycle} (a) The kiln thermal cycle for the sintering of soda-lime glass beads onto solid glass plates. (b) A close-up image of a cross section of the sintered glass beads (300--400\;\si{\um}), including a magnified region where arrows indicate the sintered interfaces between the beads.}
\end{figure*}

The kiln's thermal cycle used for the sintering process, illustrated in Fig.~\ref{fig:kiln_thermal_cycle}(a), consists of four phases, with the bead-filled glass mould being in the kiln before the start of the cycle. First, the temperature increases linearly, reaching \SI{500}{\celsius} in 3.2 hours. This is followed by a further linear increase from \SI{500}{\celsius} to \SI{700}{\celsius} over a period of three hours. The temperature is then maintained at \SI{700}{\celsius} for 2 hours to allow sintering before the final cooling phase begins. During this phase, the kiln cools naturally, leading to an exponential temperature decay (Newtonian cooling) until it reaches \SI{50}{\celsius}, when it is safe to open the kiln and take out the glass mould with the sintered beads. Here, the natural cooling was found to be sufficiently slow to prevent thermal shocks that could fracture the medium. Porous plates of different bead size were fabricated following the same protocol. The thermal cycle was established through trial and error, varying the maximum temperature and its duration, taking into account that soda–lime glass fuses at around $\SI{700}{\celsius}$ and requires slow temperature changes to avoid thermal-shock fractures \citep{Wang2013, Malou2013}. 

A photograph of a cross section of the resulted porous plate (300--400\;\si{\um}) is shown in Fig.~\ref{fig:kiln_thermal_cycle}(b). Here, the porous plate was cut with a rotating diamond tile saw, which caused some beads to detach from the cut plane and the top surface to appear more uneven than it actually is. The beads appear to have been sintered homogeneously far from the bottom glass boundary, but on the boundary we observe higher concentration of smaller beads and more compacted porous medium. Since the glass beads have a size distribution it is possible that while handling them in the mould before sintering (including shaking and vibrating), the bigger particles will stay higher while the smaller ones with go downwards and towards the bottom glass boundary \citep{Rosato1987}. This non-trivial variation in porosity along the depth of the plate may have contributed to the complexity of calibrating the light signal from imaging (see Appendix~\ref{sec:camera_calibration}). The right-hand panel of Fig.~\ref{fig:kiln_thermal_cycle}(b), which shows a magnified region, aims to highlight the sintered interfaces between the glass beads (indicated by the arrows).

\section{Porous plate permeability measurement} \label{sec:permeability_measurement}

\begin{figure}[!ht]
\includegraphics[width=0.35\textwidth]{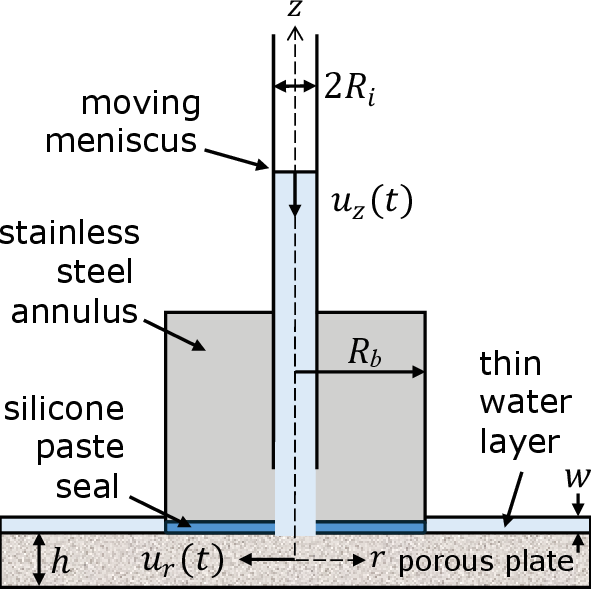}
\caption{\label{fig:permeabiity_tool} Schematics of the probe used for permeability measurements of the porous plates.}
\end{figure}

To measure the permeability of our in-house fabricated porous plates, we developed a probe consisting of a transparent plastic tube attached to an annular cylindrical stainless steel base as shown in Fig.~\ref{fig:permeabiity_tool}. Silicone moulding paste is used to seal this probe onto the surface of the porous plate. During measurements, the porous plate is maintained fully saturated as its surface is covered with a thin layer of water of depth $w \sim 5$ mm. The permeability of the medium is derived from the measurement of the height $H(t)$ of the descending water meniscus in the tube.

We assume that the porous plate of depth $\thick$ is homogenous and isotropic, neglecting any differences in the packing of the beads close to the bottom boundary. We adopt a cylindrical coordinate system with the $z$ axis ($\mathbf{\hat{z}}$) oriented vertically upward, normal to the plate and in the direction of the central axis of the cylindrical tube. Noting that $w \ (\sim 5$ mm), $\thick$ (6 mm), $R_i$ (5 mm) $ \ll R_b$ (28.6 mm), we can assume that the velocity within the tube is purely vertical, $u_z(t)$, while beneath the stainless steel annulus the velocity is purely radial, $u_r(t)$. The side glass boundary of the porous plate was always at a distance $2R_b$ or more from the central axis of the probe during the measurement. 

Darcy's law for an incompressible flow of small interstitial Reynolds number in a homogeneous and isotropic porous domain is
\begin{equation} \label{eq:darcy_law}
    \mathbf{u_D} = -\frac{\permeability}{\mu}(\boldsymbol\nabla p - \rho g \mathbf{\hat{z}}),
\end{equation}
where $\mathbf{u_D}$ is the bulk-volume-averaged velocity of the fluid, $\mu$ is the dynamic viscosity and $\rho$ the density of the fluid.

The incompressibility condition for purely radial flow,
\begin{equation} \label{eq:incompressibility_r}
\frac{\partial}{\partial r}\left(ru_r\right) = 0,
\end{equation}
combined with Eq.~(\ref{eq:darcy_law}) gives
\begin{equation} \label{eq:r_equation}
    \frac{\permeability}{\mu}\frac{\partial}{\partial r}\left(r\frac{\partial p}{\partial r}\right) = 0.
\end{equation}

Applying appropriate boundary conditions, $p=\rho g H(t)$ at $r=R_i$ and $p=0$ at $r=R_b$, and solving Eq.~(\ref{eq:r_equation}) gives {the non-hydrostatic part of the pressure field within the layer below the annulus},
\begin{equation} \label{eq:pressure(r)}
    p(r,t) = \frac{\rho g H(t)}{\ln(R_i/R_b)}\ln\frac{r}{R_b}.
\end{equation}

Mass conservation between water flowing within the column and water flowing within the porous medium under the annulus requires
\begin{equation} \label{eq:mass_conservation}
\pi R_i^2 u_z = 2 \pi h r u_r.
\end{equation}
By substituting Eq.~(\ref{eq:pressure(r)}) into Eq.~(\ref{eq:darcy_law}), we solve Eq.~(\ref{eq:mass_conservation}) for $u_z$:
\begin{equation} \label{eq:uz}
    u_z = \frac{2hr}{R_i^2}\left(-\frac{\permeability}{\mu}\frac{\rho g H(t)}{\ln(R_i/R_b)}\frac{1}{r}\right).
\end{equation}
Because $u_z(t)=-dH(t)/dt$, Eq.~(\ref{eq:uz}) can then be solved for $H(t)$, resulting in
\begin{equation} \label{eq:H}
    \ln\left(\frac{H(t=0)}{H(t)}\right) = \frac{2\permeability \rho g h t}{\mu R_i^2\ln(R_b/R_i)}.
\end{equation}
Finally, solving Eq.~(\ref{eq:H}) for permeability gives:
\begin{equation} \label{eq:kappa}
    \permeability = \frac{\mu R_i^2 \ln\left(\frac{R_b}{R_i}\right) \ln\left(\frac{H(t=0)}{H(t)}\right)}{2 \rho g h t}.
\end{equation}

Determining the time $t$ needed for the water meniscus to recede from $H(t=0)$ to some height $H(t)$ and substituting these in Eq.~(\ref{eq:kappa}) then gives the porous medium permeability $\permeability$. In applying Eq.~(\ref{eq:kappa}) it is important to note that as the dynamic viscosity of water, $\mu$, changes non-trivially with temperature, the temperature of the water used in these measurements is monitored and taken into account in the estimation of the permeability.

Some inconsistencies were observed during measurements of $\permeability$, such as an upward or downward trend in measurements performed in a row or different values on different days. Assuming that these inconsistencies resulted from gas trapped inside the pore channels of the medium, we developed a protocol that improves repeatability and smooths out steep upward or downward permeability trends. First, we saturate the porous plate with hot water (around \SI{60}{\celsius}) as, at equilibrium, hot water contains less dissolved gas than cold water. When the system is cooled down, we expect the interstitial water within the pores to be undersaturated with gas allowing any residual bubbles to be dissolved. The measurements are then conducted using de-aerated water. The permeability of the medium is calculated by averaging all the permeability values measured both on the same and on different days. We also performed confirmatory measurements at different locations across a plate and using different plates with the same nominal construction to determine the consistency of our media.

\section{Dye attenuation in a porous medium} 
\label{sec:camera_calibration}

The widely used technique of dye attenuation \citep{Holford1996,Hacker1996,Cenedese1998,Allgayer2012} relies on the intensity $I$ of a light ray of a specified wavelength $\lambda$ traversing a distance $s$ through a fluid with dye concentration $C(s)$ to follow
\begin{equation}
    \frac{dI}{ds} = -f(C)  I,
\end{equation}
where $f(C)$ is the rate of attenuation. This expression is valid only in the absence of emission through fluorescence or other mechanisms at wavelength $\lambda$. To be of value for determining $C$, we need to have knowledge of the paths through the medium taken by the light rays that reach the camera and that $f(C)$ is not affected by either the intensity $I$ (e.g., no photobleaching occurs) or chemical reactions within the flow. In the limit of low concentrations, we expect $f(C)\to\gamma C$, where $\gamma$ is a constant, recovering the classical Beer--Lambert law. This can then be integrated for a simple medium to obtain
\begin{equation}
    \frac{I}{I_0} = \exp(-\gamma\overline{C}\thick),
    \label{eq:IntegratedBeerLambert}
\end{equation}
where $\thick$ is the thickness of the medium, $\overline{C}=\frac{1}{\thick}\int_0^\thick C(s)\, ds$ is the mean concentration encountered by the light ray along its path and $I_0$ the intensity of the illumination passing through the medium when $C=0$. When $\thick$ is known, this expression may be inverted to obtain $\overline{C}$. Using imaging in the $z$ direction and a steady light source, one can then readily obtain a map of $\overline{C}(x,y,t)$ provided the rays take well-defined paths through the medium. Even if the conditions for the Beer--Lambert law are not fully satisfied (e.g., the light source or dye do not have a monochromatic response or $f(C)\ne \gamma C$), a relatively simple calibration procedure can be adopted to obtain high-quality diagnostics for $\overline{C}(x,y,t)$ \citep{Holford1996,Cenedese1998}. In applying this, it is important to remember that the digitised intensity from the camera pixels $P$ is seldom related to intensity $I$ simply via a multiplicative constant (which would disappear when evaluating $I/I_0$). Fortunately, with most modern semiconductor cameras, $P$ and $I$ are linearly related to a good approximation by $I = \beta(P - P_{black})$ across $P_{black}<P<P_{max}$. Here, $P_{black}$ is the digitised pixel values associated with no light reaching the sensor, $P_{max}$ is the maximum digitised value and $\beta$ is an unimportant constant that simply converts between the units of $P$ and those of $I$.

Despite the successful application of dye attenuation visualisation in other porous systems before \citep{Wang2008}, our work revealed additional complexities preventing a single calibration to be applied to all our experiments. The principal cause of the challenges is the refractive index contrast between the glass medium (with refractive index $n_g\approx 1.52$) and water/dye solution (with refractive index $n_s \approx 1.33$). This contrast means light rays undergo geometric scattering, encountering very large numbers of reflective and refractive fluid/glass interfaces, with only a subset of rays passing all the way through the medium. (While diffraction may play a role for very small particles, it is likely to be much less important here.) The total distance travelled by a light ray now exceeds $\int_0^\thick dz=\thick$ although, if we ignore back scattering for the moment, the distance travelled can be represented as $\int_0^\thick \frac{ds}{dz}\,dz$ where $s$ is along the ray. We expect $\frac{ds}{dz}$ to increase as the particle size is decreased due to an increase in the number of scattering events a ray undergoes. Moreover, rays perceived by a camera as exiting any particular point on the surface of the medium may have had very different paths through the porous medium (especially when originating from a diffuse light source), passing through different numbers of fluid/glass interfaces at different angles. We thus introduce the dimensionless variable $\dsdz$ such that the mean path length through pore space taken by the light rays reaching a given pixel is given by $\int_0^h \dsdz \por \,dz$. Hence, the mean total length, including the passage through glass, is approximately $\int_0^h \dsdz \,dz = \overline{\dsdz} \thick$.

To understand the impact of the scattering and porosity, we recast (Eq.~(\ref{eq:IntegratedBeerLambert}) as
\begin{equation}
    \frac{\tilde{I}}{\tilde{I}_0} = \exp\left(-\gamma\int_0^\thick \dsdz\por C \,dz\right),
    \label{eq:ComplexLambertBeer}
\end{equation}
where $\tilde{I}$ and $\tilde{I}_0$ are coarse-grained versions of $I$ and $I_0$, respectively. (Here, the coarse graining $I \to \tilde{I}$ is tailored to preserve the intensity variation of interest while removing small-scale intensity variations due to the pore structure.) We then separate the terms in the integrand into their mean and fluctuating components, $\dsdz = \overline\dsdz + \dsdz'$, $\por = \overline\por + \por'$ and $C = \overline{C} + C'$, where the over bar indicates the simple depth average $\overline{q} = \frac{1}{\thick}\int_0^\thick q\,dz$. This subdivision then gives
\begin{equation}
    \frac{\tilde{I}}{\tilde{I}_1} = \exp\left(-\gamma(\overline\dsdz\;\overline\por\;\overline{C} + \overline\dsdz\;\overline{\por' C'} + \overline\por\;\overline{\dsdz' C'} + \overline{C}\;\overline{\dsdz'\por'})\thick \right).
    \label{eq:PorousLambertBeer}
\end{equation}
The first term ($\overline\dsdz\;\overline\por\;\overline{C}$, which only vanishes when $C\equiv 0$) is equivalent to the classical result when $\overline{\dsdz}=\overline{\por}=1$. For a porous medium, the product $\overline{\dsdz}\,\overline{\por}h$ represents the mean length of pore space traversed by the rays. The other terms, however, complicate the application of (Eq.~(\ref{eq:PorousLambertBeer}). We note that each of those will vanish if both the porosity and concentration are uniform in depth (i.e., $\por' = C' = 0$). Indeed, calibration experiments (not shown here) undertaken with loose-packed glass ballotini with their pore space containing a uniform concentration of dye and illuminated with a narrow-band diffuse light source demonstrate that $\tilde{I}/\tilde{I}_1 = \exp\left(-\gamma\overline\dsdz\;\overline\por\;\overline{C}\right)$ holds to a very good approximation but, as expected, $\overline\dsdz$ depends on the particle size $\dia_p$.

For our present experiments, we note that $\por$ is approximately uniform in depth (thus we take $\por'=0$) and that additional calibration experiments and modelling of the correlation between the depth dependence of $\dsdz$ and $C$ (not included here) suggests that the contribution from temporal changes in $\overline{\dsdz'C'}$ is modest and changes little after the initial rapid surface flushing of Stage I. We hence simplify Eq. (\ref{eq:PorousLambertBeer}) and write
\begin{equation}
    \overline{C}\thick 
    = -\frac{\ln\left(\tilde{I}/\tilde{I}_0\right)}{\gamma\overline{\por}\,\overline{\dsdz}\left(1 + \overline{\dsdz'C'}/(\overline{\dsdz}\,\overline{C})\right)} 
    \approx -\scalefact\ln\left(\tilde{I}/\tilde{I}_0\right),
    \label{eq:MeanConc_Uniform}
\end{equation}  
approximating $S$ as a constant that can be determined from a single-point calibration for a given experiment.

\section{Image-based measurements} 
\label{sec:image_based_measurements}

\subsection{Dye-attenuation tracer removal against time}
\label{sec:dye_attenuation_results}

\begin{figure*}[!ht]
\includegraphics[width=0.95\textwidth]{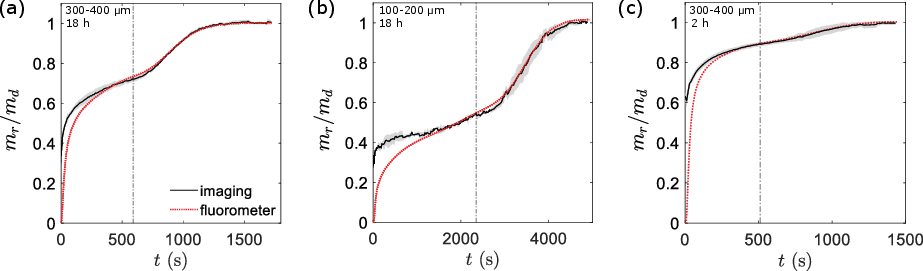}
\caption{\label{fig:imaging_vs_fluorometer} Tracer mass removed from the plate against time from imaging data of the porous plate with a dye attenuation system (black solid line). The $m_r/m_d$ curve from the effluent measurements (fluorometer data) are superimposed (red dotted line). The dotted--dashed line indicates the time of the inflection point, when the front of the dye patch reaches the end of the plate. The experimental conditions are $m_d$ = 0.2 mg, $\alpha$ = 11.0 degrees, (a) coarse plate and $\tau=$ \SI{18}{\hour}, (b) fine plate and  $\tau=$ \SI{18}{\hour}, and (c) coarse plate and  $\tau=$ \SI{2}{\hour}. The scaling factor $\scalefact$, elected for providing the best match between the two curves in each case is, $\scalefact = \SI{2.54e-7}{\gram / \milli \metre \squared}$, \SI{2.77e-7}{\gram / \milli \metre \squared}, and \SI{2.08e-7}{\gram / \milli \metre \squared}, respectively. }
\end{figure*}

Based on our imaging, the tracer removed from the plate is $m_d-m_{BL}$, where $m_{BL}$ is obtained from Eq. (\ref{eq:mass_BL}). In Fig.~\ref{fig:imaging_vs_fluorometer} this quantity non-dimensionalised with the deposited mass, $m_d$, is plotted against time (black solid line), while the respective $m_r/m_d$ curve from the effluent measurements (fluorometer data) are superimposed (red dotted line). In all the three plots of this figure $m_d$ = 0.2 mg and $\alpha$ = 11.0 degrees, while in (a) the coarse plate was used and $\tau$ = 18 h, in (b) the fine plate was used and $\tau$ = 18 h and in (c) the coarse plate was used and $\tau$ = 2 h.

The scaling factor $\scalefact$ is selected for providing the best visual match between the two curves in each case. We find that $\scalefact = \SI{2.54e-7}{\gram / \milli \metre \squared}$, \SI{2.77e-7}{\gram / \milli \metre \squared}, and \SI{2.08e-7}{\gram / \milli \metre \squared} for the plots in panel (a), (b) and (c), respectively. We observe that changing experimental conditions, such as the porous plate permeability (in panel (b)) or the dwell period (in panel (c)) has a non-negligible effect on factor $\scalefact$. Variations in the concentration measurements for experiments performed under the same conditions are small, as indicated by the standard deviation shown by the shaded range around the black curve in Fig.~\ref{fig:imaging_vs_fluorometer}(a) and (b). 

Importantly, we see that there is a very close agreement in the mass removed between the fluorometer and dye attenuation measurements beyond the rapid Stage I flushing. We note that frame rate selected for our imaging (one image every 10 or 20 seconds) means that the tracer mass reduction during the rapid flushing stage (Stage I) is not fully resolved in time. This low frame rate is not an inherent limitation of our imaging system but was selected to minimise the photo degradation of the dye that would otherwise have contaminated the fluorometer measurements.

\subsection{Image-based parametric measurements} \label{sec:parametric_measurements}

\subsubsection{Inclination angle}

\begin{figure}[!ht]
\includegraphics[width=0.8\textwidth]{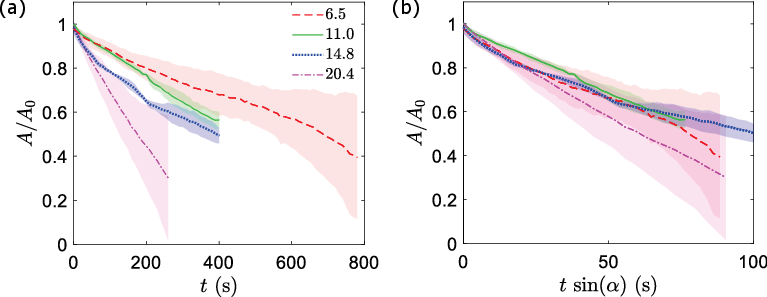}
\caption{\label{fig:area_various_angles} (a) Normalised tracer patch area against time. (b) Same as (a) but time rescaled with $\sin(\alpha)$. Experiments E1, E2, E3, and E4.}
\end{figure}

By analysing the tracer plan-view area over time ($t<t_b$) for various inclination angles, we can observe in Fig.~\ref{fig:area_various_angles}(a) that steeper inclinations lead to a faster decrease in area. This demonstrates that the increased removal rate observed in Fig.~\ref{fig:parametric_analysis}(a) for steeper $\alpha$ is not attributable to a potentially larger interfacial area between the porous medium and the clean washing film. In fact, even if $A$ was found to be larger for steeper $\alpha$ this would not significantly affect the removal rate, since the rate-limiting step is the transport within the porous medium towards the surface rather than the transfer across the porous medium--washing film interface.

\begin{figure}[!ht]
\includegraphics[width=0.8\textwidth]{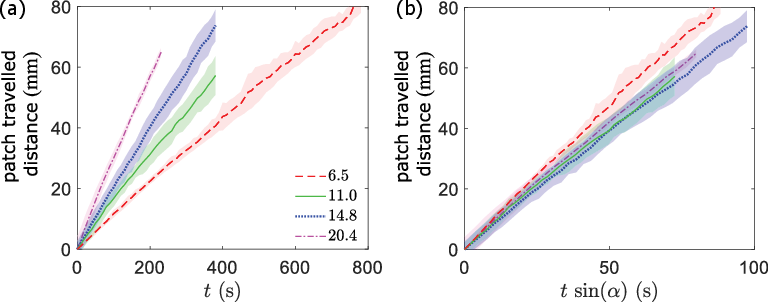}
\caption{\label{fig:patch_travel_various_angles} (a) Patch travelled distance against time. (b) Same as (a) but time rescaled with $\sin(\alpha)$. Experiments E1, E2, E3, and E4.}
\end{figure}

While steeper inclinations increase the interstitial velocity (see Fig.\ref{fig:patch_travel_various_angles}(a)) and hence longitudinal stretching of the patch via hydrodynamic dispersion, the patch area still decreases more rapidly for higher $\alpha$. Despite substantial variability in $A$, Fig.~\ref{fig:area_various_angles}(b) suggests that the evolution of $A$ for different $\alpha$ approximately collapses when time is scaled by $\sin(\alpha)$, a scaling that also applies to interstitial velocity (for a given permeability medium; see Fig. \ref{fig:patch_travel_various_angles}(b)) and removal rate (see inset of Fig.~\ref{fig:parametric_analysis}(a)). This indicates that steeper inclinations enhance not only longitudinal dispersion, causing patch elongation, but also the surface-normal transverse dispersion that promotes tracer transport towards the surface and hence increases removal rates.

\subsubsection{Permeability}

\begin{figure}[!ht]
\includegraphics[width=0.7\textwidth]{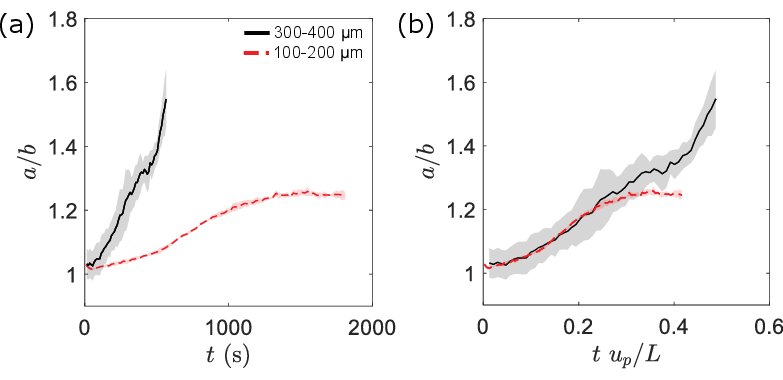}
\caption{\label{fig:ellipse_ratio_two_plates} (a) The ratio of major to minor axis of the patch fitted ellipse against time for $t<t_b$. (b) Same as (a) but time rescaled with $u_p/L$. Black solid line corresponds to the high permeability plate (glass bead size: 300-\SI{400}{\micro \metre}) and red dashed line corresponds to the low permeability plate (glass bead size: 100-\SI{200}{\micro \metre}). The rest of the conditions are identical. Experiments E7 and E8.}
\end{figure}

As longitudinal dispersion scales with the advection velocity within the porous plate, the rate of transport due to dispersion is less pronounced in the lower permeability for $t<t_b$. The patch ratio of major to minor axis of the fitted ellipse, a measure of transport due to dispersion, can be seen in Fig.~\ref{fig:ellipse_ratio_two_plates} for the coarse and fine plates, under otherwise identical conditions (experiments E7 and E8). The relative dispersion is only a function of the distance travelled, thus rescaling time with the advective time scale within the pores, $u_p/L$, causes the two curves to collapse. Although not visible in this figure, the same applies to the surface-normal component of the dispersion (transverse dispersion) that is primarily responsible for the flux of tracer into the film.

\bibliography{export}

\end{document}